\documentclass[%
 aip,
 amsmath,amssymb,
 preprint,%
 ]{revtex4-1}

\usepackage[bookmarks=true,
            bookmarksopen=true,
            bookmarksnumbered=true,
            colorlinks=true,
            linkcolor=blue,
            citecolor=blue,
            urlcolor=blue]{hyperref}

\usepackage{graphicx}
\usepackage{dcolumn}
\usepackage{bm}

\usepackage[utf8]{inputenc}
\usepackage[T1]{fontenc}
\usepackage{mathptmx}
\usepackage{etoolbox}
\usepackage{enumitem}
\usepackage{cool}
\usepackage{textgreek}

\newcommand{\JacobiCNSq}[2]{%
\JacobiCNSymb^{2}\COOL@decide@paren%
{JacobiCN}{#1 \left| \, #2 \right.\!\!}%
}

\newcommand{\JacobiDNSq}[2]{%
\JacobiDNSymb^{2}\COOL@decide@paren%
{JacobiDN}{#1 \left| \, #2 \right.\!\!}%
}

\makeatletter
\def\@email#1#2{%
 \endgroup
 \patchcmd{\titleblock@produce}
  {\frontmatter@RRAPformat}
  {\frontmatter@RRAPformat{\produce@RRAP{*#1\href{mailto:#2}{#2}}}\frontmatter@RRAPformat}
  {}{}
}%
\makeatother
\begin{document}

\preprint{AIP/123-QED}

\title[Z Pinch Kinetics II]{Z Pinch Kinetics II - A Continuum Perspective: Betatron Heating and Self-Generation of Sheared Flows}
\author{Daniel W. Crews}
\email{daniel.crews@zap.energy}
\affiliation{Zap Energy Inc., Everett, WA 98203, USA}
\author{Eric T. Meier}
\affiliation{Zap Energy Inc., Everett, WA 98203, USA}
\author{Uri Shumlak}
\affiliation{Zap Energy Inc., Everett, WA 98203, USA}
\affiliation{University of Washington, Seattle, WA 98195, USA}

\date{\today}

\begin{abstract}
  Adiabatic compression of a self-magnetizing current filament (a Z pinch) is analyzed via the adiabatic invariants of its
  constituent cyclotron and betatron motions.
  Chew-Goldberger-Low (CGL) models are recovered for both trajectories but with 
  distinct anisotropy axes, about the magnetic field for cyclotron fluid and about the
  electric current for betatron fluid.
  In particular, betatron heating produces agyrotropic anisotropy which balances with gyrophase mixing.
  A hybrid CGL model is proposed based on the local densities of cyclotron and betatron orbits,
  then validated by numerical experiments.
  The relation between anisotropy and shear is explored
  by constructing the kinetic equilibrium of a flow expanded in the flux function.
  Flow as a linear flux function is simply bi-Maxwellian, while higher powers display
  higher-moment deviations.
  Next, weakly collisional gyroviscosity (magnetized pressure-strain)
  is considered in a forward process (forced flow) and an inverse process (forced anisotropy).
  The forward process phase-mixes flow into a simple flux function, freezing flow into flux and inducing anisotropy.
  In the inverse process, betatron heating-induced anisotropy 
  self-generates a sheared flow to resist changes in flux.
  This flow, arising from momentum diffusion, is concentrated in the betatron region.
\end{abstract}

\maketitle

\section{\label{sec:intro}Introduction}
The Z pinch is a quasineutral cylindrical filament of electric current conducted through a plasma,
concentrating equilibrium plasma pressure energy in proportion to the self-magnetic energy of the current.
The Z-pinch equilibrium was originally identified in 1934 by Willard H.~Bennett, with the assistance of Llewellyn Thomas,
as a hypothesis to explain the high-voltage breakdown of gases~\cite{bennett_1934},
and termed ``the magnetic pinch effect'' in 1937 by Lewi Tonks~\cite{tonks_1937}.
The pinch as a fusion energy concept was extensively studied in the early,
classified stage of fusion research~\cite{artsimovich_1962, bostick_1977}.
Many works on pinch theory appeared at the time of fusion research declassification in 1958,
at which time Lawson~\cite{lawson_1958} and Budker~\cite{budker_parameter} independently
explained how the concept of Bennett's pinch also describes magnetically self-focused charged particle beams.
A critical parameter of magnetic self-focusing was identified,
now called the Budker parameter, given by~\cite{linhart_1959}
\begin{equation}\label{eq:budker_parameter}
  \mathfrak{B}_s = N_sr_s = \frac{v_{ts}^2}{v_s^2} = \frac{I_s}{I_{A,s}} = \frac{r_p^2}{4r_{L,s}^2} = \frac{r_p^2}{4\lambda_s^2} =
  \frac{(\omega_{cs}\tau_s)^2}{4} = \sigma_s\mathrm{Ha}^2.
\end{equation}
The manifold nature of Eq.~\ref{eq:budker_parameter} is emphasized, as various authors use the same parameter under different names.
Here $N_s$ represents particles of species $s$ per unit length, $r_s$ the classical particle radius,
$v_{ts}$ species thermal velocity, $v_s$ species partial drift velocity, $I_s$ species partial current,
$I_{A,s}$ species Alfv\'{e}n limiting current~\cite{alfven_1939},
$r_{L,s}$ Larmor radius, $\lambda_s$ skin depth, $r_p$ the pinch radius,
$\omega_{cs}$ species cyclotron frequency, $\tau_s = r_p/v_{ts}$ species thermal crossing time,
$\mathrm{Ha}$ the Hartmann number of the flow pinch, and the factors $\sigma_s$ are $\sigma_e=\sqrt{m_i/m_e}$ and $\sigma_i=\sigma_e^{-1}$.

Additional key parameters of charge neutralization fraction $f_n$ and
relativistic $\gamma$-factor were noted in 1959 by Lawson~\cite{lawson_1959}.
The relations of Eq.~\ref{eq:budker_parameter} involving species velocity
hold for both electrons and ions in the Lorentz frame where the two are counter-streaming
and thus perfectly charge neutralized (for $f_n=1$).
The relations of Eq.~\ref{eq:budker_parameter} are derived with commentary in the prequel~\cite{kinetics_i}.

The Budker parameter 
measures the ratios of: temperature to drift-kinetic energy,
plasma current to Alfv\'{e}n limiting current, characteristic Larmor radius/skin depth to pinch radius,
the species dynamical Hall parameter, and the magnitude of electromagnetic to viscoresistive hydrodynamic forces.
Perhaps most importantly, $\mathfrak{B}_s$ measures the magnetization of orbits.
When $\mathfrak{B}_s\gtrsim 1$, a non-negligible part of the canonical ensemble form cyclotron orbits,
initially in the periphery where magnetic flux density is high.
The remaining particles in the current-dense core follow betatron orbits, a radially bouncing, axis-encircling drift motion associated with
magnetically self-focused but ``unmagnetized'' streams. 
The prequel~\cite{kinetics_i} shows that the ensemble-averaged canonical momentum $P_z$ passes from an unbound state $P_z>0$
to a bound state $P_z<0$ through $\mathfrak{B}_s=1$, analogous to the positive and negative energies of gravitational or electrostatic binding~\cite{arnold_2013, kinetics_i}.

The limit $\mathfrak{B}_s\to 0$ is the negligible self-field neutralized beam limit (typically with $\gamma \gg 1$),
while $\mathfrak{B}_s\to \infty$ is the magnetohydrodynamic (MHD) limit~\cite{finkelstein_1959}.
For neutralized beams, MHD modes are thought to be stabilized by 
betatron orbit effects~\cite{bennett_1955, finkelstein_1959},
with a residual stabilizing effect when $\mathfrak{B}_i>1$ in the
``large ion Larmor radius'' (LLR) regime~\cite{HAINES_2001} ($\mathfrak{B}_i\approx\mathcal{O}(10)$).
Interestingly, the terms ``LLR Z pinch'' and ``neutralized ion beam above the Alfv\'{e}n limit''
are synonymous in the perfectly neutral Lorentz frame but not synonymous in the laboratory frame,
distinguished by the laboratory-frame axial velocity of the ion betatron orbits.
The term ``neutralized beam above the Alfv\'{e}n limit'' is usually applied to electrons
in the relativistic regime~\cite{santos_2018}.
Betatron resonance is receiving increasing attention in the context of plasmas forced by lasers near the critical density~\cite{arefiev_2024}.

The frame-dependent syzygy of neutralized ion beams and LLR Z pinches in the $\mathfrak{B}_i\approx 1$ regime is closely related
to the notorious Z-pinch beam-target fusion mechanism~\cite{vikhrev_2007}.
Experimental Thomson scattering measurements on the Fusion Z-Pinch Experiment (FuZE)~\cite{goyon_2024}
suggest that the deuterium Budker parameter $\mathfrak{B}_d\approx 20$ at the time of neutron radiation,
while for the electrons $\mathfrak{B}_e\approx 7\times 10^{4}$.
Neutron isotropy measurements on FuZE have been found to be inconsistent with typical Z-pinch beam-target
fusion mechanisms~\cite{mitrani_2021}, suggesting LLR Z-pinch laboratory-frame ion dynamics
and thereby motivating the present study of kinetic effects in flow pinches. 


In the prequel, the distribution of orbits was studied in the transitional magnetization regime
(\textit{i.e.}, moderate ion Budker parameters, or the LLR regime).
It was found that plasma species density is decomposable into 
betatron and cyclotron orbital densities, $n_s=n_{\beta,s}+n_{c,s}$.
The net density $n_s$ is fully composed of betatron orbits as $r\to 0$, fully of cyclotron orbits as $r\to\infty$,
and partially of both at intermediate radii within a transition layer through which
current conduction switches from betatron to diamagnetic cyclotron conduction.
By considering the canonical distribution (the Bennett pinch), the location and thickness of
this transition layer relative to the pinch radius was found to depend only on $\mathfrak{B}_s$,
and lies close to the pinch radius in the LLR regime.
This kinetic description of the Bennett pinch is a diffuse example of Kotelnikov's theory of sharp
diamagnetic layers in the Gas Dynamic Trap~\cite{Kotelnikov_2020}.
Some current conduction by betatron orbits appears to be a necessary element
of $\beta\geq 1$ plasma configurations~\cite{timofeev_2024}.

In this second work on Z-pinch kinetics, we apply the results of the prequel to the topics of
collisionless adiabatic compression, kinetic equilibrium, and self-organization of sheared flows in the Z pinch.
Collisionless momentum and heat fluxes in the intermediate magnetization regime are of particular
interest to flow Z-pinch dynamics because Braginskii-type closures~\cite{braginskii1965} 
do not describe the betatron-orbiting part of the plasma column even on moderately collisional timescales.
On the weakly collisional timescales, phase-space mixing produces effective dissipation.
While this allows hydrodynamic intuition to be applied somewhat to collisionless dynamics~\cite{senbei_2020, bandyopadhyay_2023},
kinetic effects can lead to unexpected phenomena such as flow-organizing and flow-generating viscosity~\cite{del_sarto_aniso, del_sarto_how_anisotropization}.

Collisionless momentum fluxes are often studied in the context of the magnetized
pressure-strain (or Pi-D) interaction~\cite{yang_2017},
coinciding in the collisional limit with gyroviscosity~\cite{kaufman_1960}.
The collisionless viscosity associated with magnetized shear flows is observed to be an important element
of astrophysical plasma turbulence at the ion scale~\cite{hellinger_2024},
for example in the magnetosphere~\cite{bandyopadhyay_2020} and solar wind~\cite{pezzi_2019}.
A thorough discussion of the pressure-strain interaction can be found in a series of articles
by Cassak and Barbhuiya et al~\cite{cassak_2022_1, cassak_2022_2, hasan_2022}.
Because LLR Z pinches are themselves on the ion inertial scale, flow pinch physics
naturally parallels the astrophysical studies.
Notably, the phase space signature of the unmagnetized pressure-strain interaction has recently been studied
and found to be closely related to the field-particle correlation~\cite{conley_2024}, suggesting a fruitful line of
investigation for similar relationships in the kinetic theory of magnetized flows.

This work makes several novel contributions to Z-pinch physics which should be of interest to the general fusion community, including:
\begin{itemize}[topsep=0pt,itemsep=-1ex,partopsep=1ex,parsep=1ex]
  \item An adiabatic model for anisotropic response to rapid fluctuations,
    applicable to both the cyclotron and betatron fluids (Section~\ref{sec:collisionless_compression},
    subsequently validated in Section~\ref{sec:numerical_validation}),
  \item A class of kinetic equilibria where axial flow is a power series in the flux function,
    demonstrating increasingly non-thermal features with higher powers in the flux (Section~\ref{sec:model-distributions}),
  \item A study of dynamic kinetic gyroviscosity on weakly collisional timescales, in which gyro-phase mixing rapidly brings
    forced shear flows towards kinetic equilibrium (Section~\ref{sec:gyroviscous_damping}),
  \item Demonstration of spontaneous shear flow generation due to flux compression of the betatron fluid,
    exhibiting reactive resistance to rapid changes in magnetic flux (Section~\ref{sec:spon_sheared_flow}),
  \item An equivalence between the characteristic anisotropy of an Alfv\'{e}nic sheared flow with the Budker parameter,
    indicating increased kinetic-instability-mediated viscosity in the low ion Budker regime (Section~\ref{subsec:char_anisotropy}), 
\end{itemize}

These results are developed as follows.
First, Section~\ref{sec:kinetic-regime} considers conditions for kinetic processes faster than the thermalization time,
Section~\ref{sec:collisionless_compression} develops a model for the anisotropy produced by fast compression,
and Section~\ref{sec:model-distributions} presents a theory of kinetic flow-pinch equilibrium.
Collisionless gyroviscosity (magnetized pressure-strain interactions)
is discussed in Section~\ref{sec:flows_sec}, 
while Section~\ref{sec:numerical_validation} concludes with a numerical validation of the anisotropy model.

\section{\label{sec:kinetic-regime}Kinetic dynamics in the flow Z pinch}
The compressibility, transport properties, and stability of plasmas essentially depend
on the ion distribution function, which in turn is governed by kinetic processes.
Of particular interest then is the ion relaxation rate $\nu_{ii}$,
which is the rate at which ions approach a Maxwellian distribution from intraspecies scattering.
The ion relaxation rate is proportional to their density and inversely proportional to temperature
to the 3/2 power~\cite{braginskii1965},
\begin{equation}\label{eq:collision_timescale}
  \nu_{ii} = \frac{\omega_{pi}}{9\sqrt{\pi}}\frac{\ln\Lambda}{\Lambda_i} \sim \frac{n}{T_i^{3/2}}\ln\Lambda
\end{equation}
with $\omega_{pi}$ the plasma frequency, $\ln\Lambda$ the Coulomb logarithm, and $\Lambda_i=\frac{4}{3}\pi n_i\lambda_{Di}^3$ the ion plasma parameter.
The Z pinch may be heated by adiabatic compression to increase the ion density and temperature~\cite{shumlak_fusion2020}.
Importantly, the ion relaxation rate varies only logarithmically during compression along an adiabat,
as, if $\nu_0$ is a reference relaxation rate, one can show that
\begin{equation}\label{eq:relaxation_as_adiabatic}
  \frac{\nu_{ii}}{\nu_0} = e^{-(s-s_0)/R}\frac{\ln\Lambda}{\ln\Lambda_0}
\end{equation}
where $s = c_v\ln(p/n^\gamma)$ is specific entropy, the specific heat ratio $\gamma=5/3$ for monatomic ions,
and $c_v=R/(\gamma-1)$ is the specific heat at constant volume given the ion gas constant
$R=k_B/m_i$ for $k_B$ Boltzmann's constant.

Equation~\ref{eq:relaxation_as_adiabatic} shows that the relaxation rate is approximately constant in an adiabatic process.
The adiabatic approximation is accurate in the collisional limit, but is modified by collisionless orbits
and shock wave dynamics. 
Adiabatic processes include transient compression as induced by rising discharge current~\cite{ryutov_2000}
and steady flow compression as discussed by Morozov and Solov'ev~\cite{Morozov1980}.
Adiabatic flow-pinch compression was recently discussed by Turchi and Shumlak~\cite{turchi_zpinch},
and ideal flow-pinch thermodynamics was recently studied by the authors~\cite{crews_kadomtsev}.

Classical Z-pinch transport physics depends on the magnetization of ion orbits and the ion relaxation rate.
Ion orbit magnetization is measured by the ion Budker parameter $\mathfrak{B}_i$,
while dependence on the ion relaxation rate is measured by dimensionless parameters
including the ion Hall parameter $\chi_i \equiv \omega_{ci}/\nu_{ii}$ with $\omega_{ci}$ the characteristic cyclotron frequency,
and the ratio of the plasma dynamical frequency $\nu_d$ to the relaxation rate, $\chi_d \equiv \nu_d/\nu_{ii}$.
The dynamical frequency $\nu_d = \tau_d^{-1}$ is the inverse of the dynamical timescale $\tau_d$ of interest,
such as the plasma flow-through time (\textit{i.e.}, $\tau_d = L/v_z$ with $L$ plasma length and $v_z$ axial fluid velocity),
the flux compression timescale, or the growth rate $\gamma$ of instability modes,
which is typically estimated by a characteristic frequency
(\textit{e.g.}, Alfv\'{e}n transit rate $\nu_A = v_A/r_p$ for MHD modes or plasma frequencies for microinstabilities).

The magnetized transport theory of Braginskii~\cite{braginskii1965}, and extensions thereof~\cite{Hunana_2022},
is valid in the regime $\mathfrak{B}_i\gg 1$, $\chi_i\ll 1$, and $\chi_d\ll 1$.
Important corrections may be determined for the flow Z pinch.
For instance, as a unity-$\beta$ plasma, orbit magnetization is not radially uniform.
The experimental value of $\mathfrak{B}_i=20$ suggests that $\approx 30\%$ of the total radial ion inventory
follow betatron trajectories in a near-axis transitional magnetization layer~\cite{kinetics_i},
within which radius magnetized orbit theory is not correct.
Modification to gyroviscosity is a particularly important correction for the flow Z pinch.
Construction of a transport theory valid for radii in the transitional magnetization layer is an open problem.
Meier and Shumlak have proposed empirical adjustments to avoid singularities in the Braginskii transport coefficients around the axis~\cite{meier_two_fluid}.

When dynamical timescales are faster than the collision time ($\chi_d \gtrsim 1$) Braginskii closures
are not valid for the cyclotron orbits either because adiabatic dynamics lead to non-Maxwellian features
beyond the closure model.
In this situation, fluid elements no longer conserve specific entropy but instead
conserve multiple invariants associated with the adiabatic invariants of their constituent particles.
Multiple adiabatic invariants lead to anisotropic dynamics.

Experimentally relevant collisionless dynamics 
may arise during steady-flow compression at sufficiently high velocities or small length scales,
transient flux compression events, and MHD instability growth at large ion Hall parameter.
In addition, the consequences of these collisionless dynamics persist when the plasma lifetime is
shorter than the relaxation time.
The plasma lifetime may be considered the shorter of either the reactor flow-through time or the MHD instability growth time.
Collisional processes can also interact in feedback with collisionless dynamics, such as the case of
microinstability-enhanced resistivity~\cite{ryutov_2015}.

\begin{figure}[h!]
  \includegraphics[width=0.618\linewidth]{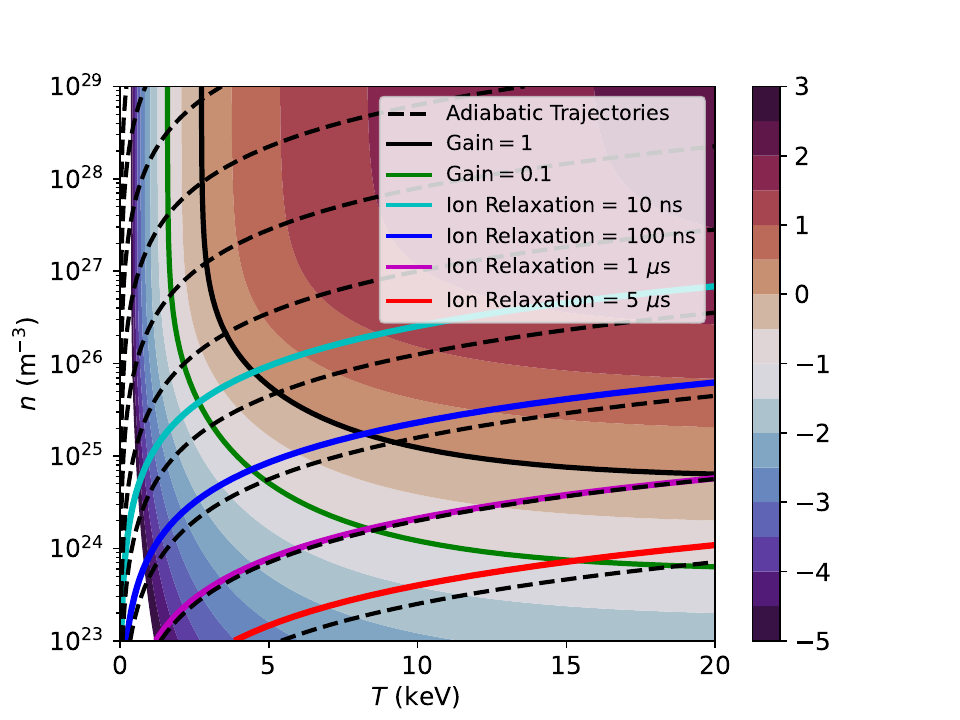}
  \caption{
    Filled contours show fusion gain~\cite{shumlakgain2024} $Q$ in temperature/density
    space $(T, n)$ for a flow Z pinch with axial velocity $v_z = 100$ km/s, length $L=50$ cm,
    and plasma flow-through time $\tau = 5$ $\mu$s.
    Black/green contours mark $Q=1$ and $0.1$, respectively.
    Dashed lines show adiabatic trajectories ($p/\rho^\gamma=\text{const.}$).
    Solid contours indicate ion relaxation times from 10 ns (cyan) to 5 $\mu$s (red).
    The breakeven region ($T\in (5,20)$ keV, $n\in (10^{24},10^{25})$ m$^{-3}$)
    has ion relaxation times comparable to the plasma flow-through time,
    suggesting potentially significant anomalous transport effects.
    \label{fig:q}}
\end{figure}

Figure~\ref{fig:q} shows the parameter space of the flow Z pinch in terms of ion temperature and density
alongside calculated fusion gain contours~\cite{shumlakgain2024}.
The filled contours show the fusion gain, while the colored lines show the ion relaxation time.
The dashed lines show adiabatic trajectories, \textit{i.e.}~paths followed by the state of the plasma
as it is heated by adiabatic compression.
The black and green lines show fusion gain near and close to break-even level, $Q=0.1$ and $Q=1$ respectively.

Figure~\ref{fig:q} shows that collisionless dynamics are relevant to breakeven fusion conditions in flow Z pinch plasmas,
in this case where the ion relaxation time is $\mathcal{O}(0.1)-\mathcal{O}(1)$ $\mu$s and thus
comparable to the plasma lifetime and dynamical scales of interest
(\textit{i.e.}, flow-through time, instability growth time, and compression time).
Importantly, the ion relaxation time on an adiabat has the significance of the maximum time
over which compression is singly adiabatic.
Compression faster than the ion relaxation time, which we refer to as collisionless flux compression,
produces anisotropic dynamics with a variety of interesting consequences.






\section{Adiabatic model of anisotropy in a compressing Z pinch\label{sec:collisionless_compression}}
Dynamical processes on collisionless timescales generate pressure anisotropy,
thus changing plasma equilibrium, momentum and heat transport, and kinetic stability~\cite{Bott_2024}.
The focus of this section is on pressure anisotropy generated by the dynamical process of collisionless flux compression.
A model is developed which accounts for the transitional magnetization of pinch orbits from the
current-dense center to the magnetized periphery, and applied in later sections to consider collisionless
modifications to cross-field viscosity.

Compression is isotropic and collisional when slower than the relaxation time (see Braginskii~\cite{braginskii_rev1965}, page 260,
for a lucid description),
and is adiabatic with respect to a single invariant, namely the specific entropy of the fluid elements.  
On the other hand, collisionless trajectories are associated with multiple adiabatic invariants,
and collisionless compression which preserves these invariants can be described by a multiply adiabatic theory.
Multiply adiabatic collisionless compression may be an inaccurate description in the presence of
non-adiabatic collisionless processes such as shocks, radiation, and resistivity.  

The set of single-particle adiabatic invariants depends on the particle orbit magnetizations.
Well-magnetized particles on cyclotron trajectories conserve their magnetic moments and their angular momentum
(or mirror invariant in a slab geometry),
leading to doubly adiabatic compression described by the Chew-Goldberger-Low (CGL) equations~\cite{cgl_closure}.
Magnetic moment conservation produces gyrotropic anisotropy, meaning the anisotropy symmetry axis is along magnetic field lines.
In contrast, the adiabatic invariants of betatron trajectories lead to agyrotropic anisotropy under compression,
meaning the anisotropy is symmetric about the electric current lines.



\subsection{The CGL equations for fluid elements of cyclotron trajectories in a cylinder\label{sec:cgl_cyclotron}}
Doubly adiabatic CGL theory describes collisionless compression of magnetized plasma,
and predicts different degrees of parallel and perpendicular heating due to magnetic flux compression~\cite{cgl_closure}.
The classic CGL model consists of two invariants, one for each pressure component
\begin{subequations}\label{eq:cgl_eqns}
\begin{eqnarray}
  \frac{d}{dt}\Big(\frac{p_\perp}{\rho B}\Big) &=& 0,\\
  \frac{d}{dt}\Big(\frac{p_\parallel B^2}{\rho^3}\Big) &=& 0,
\end{eqnarray}
\end{subequations}
in contrast to the singly adiabatic (conserved specific entropy) invariant $\frac{d}{dt}\big(\frac{p}{\rho^\gamma}\big)=0$.

Equations~\ref{eq:cgl_eqns} hold in the cylindrical Z pinch for fluid elements which are made up primarily of cyclotron trajectories.
This is demonstrated simply as follows.
Suppose that a plasma fluid element is characterized by three adiabatic invariants,
namely specific magnetic moment $\mu$, specific angular momentum $L_\theta$, and specific magnetic flux $s_m$ (from Alfv\'{e}n's law)
\begin{subequations}\label{eq:three_adiabatic_invariants}
\begin{eqnarray}
  \frac{d\mu}{dt}=0 &\quad\implies\quad& \frac{d}{dt}\Big(\frac{v_{\mathrm{th},\perp}^2}{B_\theta}\Big)=0,\\
  \frac{dL_\theta}{dt}=0 &\quad\implies\quad& \frac{d}{dt}(rv_{\mathrm{th},\parallel})=0,\\
  \frac{ds_m}{dt}=0 &\quad\implies\quad& \frac{d}{dt}\Big(\frac{B_\theta}{\rho r}\Big)=0
\end{eqnarray}
\end{subequations}
where $v_{\mathrm{th},\parallel} = \sqrt{k_BT_\parallel/m}$ is the parallel thermal velocity
and $v_{\mathrm{th},\perp}=\sqrt{k_BT_\perp/m}$ the perpendicular one.
Then substituting the parallel and perpendicular pressures
($p_\parallel = nkT_\parallel$ and $p_\perp=nkT_\perp$) for the thermal velocities
and combining Eqs.~\ref{eq:three_adiabatic_invariants} yields exactly Eqs.~\ref{eq:cgl_eqns}, the CGL equations.

Equations~\ref{eq:three_adiabatic_invariants} clarify the double-adiabatic CGL model in cylindrical geometry
as consisting of the magnetic moment plus the angular momentum invariant, versus the Cartesian picture
in which the second invariant is the mirror bounce action $J_\parallel \sim v_\parallel\ell$
along a field-aligned path of length $\ell$. 
The singly adiabatic model (invariant specific entropy) arises in the collisional limit $\omega_{ci}\tau\ll 1$.
Observe that the CGL model is truly triply adiabatic considering the necessity of Alfv\'{e}n's law to obtain the CGL equations
and the thermodynamic character of the specific magnetic flux~\cite{crews_kadomtsev}.

Compression in the Cartesian slab model and in the cylindrical pinch are distinguished by the Jacobian appearing in
Alfv\'{e}n's law~\cite{turchi_zpinch}.
Namely, the Cartesian slab has $\frac{d}{dt}\big(\frac{B}{\rho}\big) = 0$ while the cylinder has $\frac{d}{dt}\big(\frac{B_\theta}{\rho r}\big)=0$, a subtle difference which can lead to opposite expectations.
That is, combining Eqs.~\ref{eq:cgl_eqns} with these frozen flux invariants produces the following equations,
\begin{subequations}\label{eq:slab_vs_cylinder}
\begin{eqnarray}
  \textrm{Slab, }&\frac{d}{dt}\Big(\frac{p_\parallel}{p_\perp}B\Big) = 0,&\\
  \textrm{Cylinder, }&\frac{d}{dt}\Big(\frac{p_\parallel}{p_\perp}B_\theta r^2\Big) = 0.&\label{eq:cylinder_eq}
\end{eqnarray}
\end{subequations}
Cartesian slab anisotropy is simply inversely proportional to the flux compression ratio,
$\frac{p_{\parallel,2}}{p_{\perp,2}}=C_m^{-1}$ (with $C_m\equiv \frac{B_2}{B_1}$)
between an isotropic initial state $(\cdot)_1$ and an anisotropic compressed state $(\cdot)_2$.
In the same vein, anisotropy in the cylinder follows from Eq.~\ref{eq:cylinder_eq} as
\begin{equation}\label{eq:cyclotron_anisotropy}
  \frac{p_{\parallel,2}}{p_{\perp,2}} = \Big(\frac{I_1}{I_2}\Big)\Big(\frac{r_1}{r_2}\Big)
\end{equation}
where $\mu_0 I \equiv 2\pi rB_\theta$ is the axial current enclosed up to radius $r$.

Thus, the anisotropy produced by flux compression of cyclotron trajectories in a Z pinch is the
product of two factors: the current amplification ratio and the ratio of radii
(alternatively, the product of the flux and area compression ratios).
Flux compression in the Cartesian slab always leads to the anisotropy orientation $p_\parallel < p_\perp$,
but the Z pinch will easily display the orientation
$p_\parallel > p_\perp$ if current does not increase by a greater fraction than the radius.
This azimuthal over-heating can happen because angular momentum conservation competes with magnetic moment conservation,
and has important consequences for the resulting radial force balance.

\subsection{Adiabatic invariants of betatron orbits and CGL model of the betatron fluid\label{sec:cgl_betatron}}
To summarize the preceding, fluid elements composed of cyclotron trajectories display two adiabatic invariants
(Eqs.~\ref{eq:three_adiabatic_invariants}), magnetic moment and angular momentum, and an additional frozen-in flux invariant.
In the cylinder, magnetic moment is the invariant associated with radial and axial bounce of the cyclotron orbit,
and the angular momentum with periodic azimuthal orbits.

Idealized betatron trajectories display invariants closely related to those of the ideal cyclotrons
with a crucial difference: the lines of electric current are the principal symmetry axis rather than the magnetic field lines.
Ergo, fluid elements composed of betatron trajectories should follow the same CGL equations as
the cyclotrons, but the notions of parallel and perpendicular refer to the electrical current axis ($\hat{z}$) instead of
the magnetic field lines, which we demonstrate below.

The adiabatic invariants of an ideal betatron orbit consist of its radial bounce action $J_r$,
its axial bounce action $J_z$, and its azimuthal angular momentum $L_\theta$ (or action of the periodic azimuthal motion).
These invariants are inferred from the ideal betatron orbit (cf. the prequel~\cite{kinetics_i}),
which we present here for self-completeness.

The radial potential energy of a charged particle in the vicinity of the Z-pinch axis is
\begin{equation}
  U = \frac{(P_z - q_sA_z)^2}{2m_s} \approx \frac{P_z^2}{2m_s} - P_z\frac{q_sA_z}{m_s}
\end{equation}
with $P_z$ the z-component of canonical momentum, provided that $P_z \gg -\frac{1}{2}q_sA_z$.
Expanding the vector potential to $\mathcal{O}(r^2)$ around the Z-pinch axis 
yields a harmonic potential $U \approx \frac{1}{2}kr^2$ describing a radial bounce with spring constant $k$.
The axial bounce motion follows from conservation of canonical momentum.
These radial and axial bounce motions are expressed as
\begin{subequations}\label{eq:ideal_betatron_orbit}
\begin{eqnarray}
  r(t) &=& r_0\sin(\omega_\beta t),\\
  mv_z(t) &=& P_z - \kappa\sin^2(\omega_\beta t).
\end{eqnarray}
\end{subequations}
The axial momentum fluctuation amplitude $\kappa\equiv q_sA_z(r_0)$ is the potential momentum at the radial turning point $r_0$
and $\omega_\beta$ is the betatron frequency with $\omega_\beta^2\equiv \frac{q_sB_\theta}{m_s}\frac{v_z}{r}$.
The radial and axial bounce motions contribute substantially to temperature in the species drift frame.

The radial bounce invariant follows as the area of the $(r,v_r)$ phase-plane orbit,
\begin{equation}\label{eq:betatron_action}
  J_r = \pi m_s r_0^2\omega_\beta \sim v_{r0}r_0,
\end{equation}
or the product of the radial bounce amplitude $r_0$ and the radial velocity amplitude $v_{r0}$.
The radial bounce action is similar in form to the angular momentum or mirror bounce invariants,
\textit{i.e.} velocity times length, as a simple harmonic motion.

The axial bounce invariant is found by transforming to the bounce-averaged ($\langle\cdot\rangle_T$) axial drift frame,
\textit{i.e.} to the oscillation-center coordinates $(\delta z, \delta v_z)$.
The oscillation-center trajectory is
\begin{subequations}\label{eq:axial_bounce_motion}
\begin{eqnarray}
  \langle m_sv_z\rangle_T &=& P_z - \frac{\kappa}{2},\\
  \delta(m_sv_z)\equiv m_sv_z - \langle m_sv_z\rangle_T &=& -\kappa\cos(2\omega_\beta t),
\end{eqnarray}
\end{subequations}
and the axial bounce invariant as the area of the oscillation-center phase-plane ellipse,
\begin{equation}\label{eq:axial_betatron_action}
  J_z = \pi \frac{\kappa^2/2m_s}{\omega_\beta}.
\end{equation}
Equations~\ref{eq:betatron_action} and~\ref{eq:axial_betatron_action} hold approximately for non-ideal betatron orbits,
in a similar way that magnetic moment invariance is taken to hold for cyclotron orbits,
although it is truly conserved only up to guiding-center corrections.
We also consider axis-encircling betatron orbits to be approximately described by
Eqs.~\ref{eq:betatron_action} and~\ref{eq:axial_betatron_action} with the addition of an
angular momentum invariant, $J_\theta = m_sr_0v_\theta$, because betatron orbits with
azimuthal momentum may be approximated as an in-plane bounce orbit that rotates around the $\hat{z}$-axis~\cite{kinetics_i}.

We now specialize to ion trajectories because their orbits are much larger than the electrons.
The characteristic betatron frequency $\omega_\beta$ follows from $\omega_\beta^2\equiv \omega_{ci}\frac{v_z}{r}$ by
setting $v_z$ equal to the ion axial drift $v_i$ (in the perfectly neutral Lorentz frame).
Three useful forms of this frequency are  
\begin{equation}\label{eq:three_useful_betatron_frequencies}
  \omega_\beta = \frac{v_{\mathrm{th},i}}{r_p} = \frac{v_i}{\lambda_i} = \omega_{ci}\mathfrak{B}_i^{-1/2}
\end{equation}
where $v_{\mathrm{th},i}$ is ion thermal velocity, $r_p$ is pinch radius, $\lambda_i = c/\omega_{pi}$ is ion inertial length,
$\omega_{ci}$ is characteristic ion cyclotron frequency, and $\mathfrak{B}_i$ the ion Budker parameter~\cite{kinetics_i}.

Applying Eq.~\ref{eq:three_useful_betatron_frequencies} to Eq.~\ref{eq:axial_betatron_action} and
supposing $\mathfrak{B}_i=\text{const.}$, we infer three Lagrangian invariants
associated to an ion fluid element threaded by betatron orbits, in analogy to Eqs.~\ref{eq:three_adiabatic_invariants},
to be
\begin{subequations}\label{eq:three_adiabatic_betatron_invariants}
\begin{eqnarray}
  \frac{d}{dt}\Big(\frac{v_{\mathrm{th},z}^2}{B_{\theta,0}}\Big)=0,\label{eq:Jax}\\
  \frac{d}{dt}(rv_{\mathrm{th},r\theta})=0,\label{eq:Jrth}\\
  \frac{d}{dt}\Big(\frac{B_{\theta,0}}{\rho r}\Big)=0\label{eq:Jf}
\end{eqnarray}
\end{subequations}
having assumed $(r,\theta)$-isotropy,
corresponding to the thermal axial invariant (Eq.~\ref{eq:Jax}), radial invariant (Eq.~\ref{eq:Jrth})
and azimuthal angular momentum invariant (Eq.~\ref{eq:Jrth}).
In Eqs.~\ref{eq:three_adiabatic_betatron_invariants}, $B_{\theta,0}$ is the characteristic flux density
in the vicinity of the magnetic O-point about which the betatron orbits are bouncing.
Combination of Eqs.~\ref{eq:three_adiabatic_betatron_invariants} obtains the CGL model for
an ion betatron fluid element,
\begin{subequations}\label{eq:cgl_eqns_betatrons}
\begin{eqnarray}
  \frac{d}{dt}\Big(\frac{p_{r\theta}}{\rho B_{\theta,0}}\Big) &=& 0,\\
  \frac{d}{dt}\Big(\frac{p_zB_{\theta,0}^2}{\rho^3}\Big) &=& 0.
\end{eqnarray}
\end{subequations}
The electrical current axis assumes the principal anisotropy axis of the betatron fluid,
and is otherwise modeled by the same CGL equations as a cyclotron plasma fluid.
The radial and azimuthal directions display harmonic motions essentialized by
a characteristic distance and a characteristic velocity (similar to the longitudinal mirror invariant),
and the harmonic motion of the axial direction is essentialized by characteristic kinetic energy and the
characteristic cyclotron frequency (like the magnetic moment of the cyclotrons).
Anisotropic heating of the radial and axial directions is an intuitive consequence of the inductive
axial electric field associated with flux compression splitting its energy between:
1) acceleration of the betatron axial drift, and
2) particle heating through increasing the energy of the betatron's radial, axial, and azimuthal periodic motions.

Equations~\ref{eq:three_adiabatic_betatron_invariants} come with important caveats.
These invariants have been inferred based on the properties of ``linear'' betatron orbits, for which $P_z\gg -q_sA_z/2$,
which translates to a very small trapping parameter in the prequel's terminology~\cite{kinetics_i}.
This analysis based on the linear orbit's invariants parallels invariance of the magnetic moment for
the cyclotron trajectories only to lowest order in the finite Larmor radius (FLR) parameter,
\textit{viz}.~trapping parameter.
The action integrals of a particular orbit are truly nonlinear functions of its trapping parameter.

Indeed, for the Bennett distribution, nonlinearity is non-negligible for many near-axis orbits.
Sufficiently close to the magnetic O-point, these adiabatic invariants are expressed in elliptic integrals.
Such nonlinear expressions for the radial betatron action were given explicitly by Sonnerup~\cite{sonnerup_1971}
in the context of a reconnecting current sheet.
Sonnerup's analysis identified a significant fact: adiabatic compression of the center of the current sheet
causes all orbits (both cyclotrons and betatrons) to approach the same elliptic modulus (trapping parameter)
$\alpha = 2^{-1/4} \approx 0.84$ (at constant canonical momentum $P_z$).
This trapping parameter $\alpha < 1$ represents a nonlinear betatron oscillation~\cite{kinetics_i}.
Thus, a corollary of Sonnerup's result, originally identified for adiabatic compression of a reconnecting current sheet,
is that sufficient flux compression of the Z pinch drives all near-axis particles into a nonlinear betatron orbit.

Sonnerup's result assumes that the non-adiabatic cyclotron-betatron separatrix crossing is
adiabatic-in-mean, as argued by Janicke~\cite{Janicke_1975}.
In reality, the separatrix crossing is chaotic, as explored by Cary et al.~\cite{cary_1986} and B\"{u}chner~\cite{buchner_1989},
which results in compression timescale-dependent corrections to the spectrum of the resulting betatron orbit flux.
We note however that such non-adiabatic cross-separatrix orbital variations are less significant
for the compressing Z pinch than the reconnecting current sheet because the infinite-period separatrix only occurs
for purely radial/axial bounce orbits.
Non-zero angular momentum of axis-encircling orbits erases the infinite-period separatrix,
although all trajectories passing directly through $r=0$ may undergo these chaotic non-adiabatic changes.
Non-adiabatic transitions across the betatron-cyclotron separatrix of axis-crossing orbits in a compressing Z pinch
deserves deeper investigation in a future study.

\subsection{Simple model of anisotropic heating from flux compression of a Z pinch}
Sections~\ref{sec:cgl_cyclotron} and~\ref{sec:cgl_betatron} established that the CGL model applies to plasma
composed of either cyclotron or betatron trajectories, in the usual way for the cyclotron plasma and a
modified way for the betatron plasma.
Specifically, the models predict the same magnitude of anisotropic heating from flux compression of a pinched plasma
displaying either flavor of motion, but predict differing principal anisotropy axes; either the magnetic field lines
for fluid consisting of
cyclotron trajectories, or the electrical current lines for a fluid of betatron trajectories.

We now combine Sections~\ref{sec:cgl_cyclotron} and~\ref{sec:cgl_betatron}
to propose an anisotropy model for a compressing Z pinch plasma, similar to the Ochs-Fisch model~\cite{ochs_anisotropy}
but valid for all radii.
The model consists of estimating a radially uniform magnitude of anisotropy due to flux compression using the CGL model
and calculating the principal axis of the anisotropic Maxwellian thus generated in terms of the local partial pressures of
cyclotron and betatron trajectories.
Based on numerical experiments (Section X), this model appears accurate under the assumptions:
\begin{enumerate}[topsep=0pt,itemsep=-1ex,partopsep=1ex,parsep=1ex]
\item The flux function varies slowly compared to the Alfv\'{e}n transit time, and thus changes uniformly throughout the plasma;
\item The change in particle magnetization during compression (due to separatrix crossing)
  is small enough that the relative densities of cyclotron and betatron trajectories are invariant.
\end{enumerate}
Assumption 1 means that Eq.~\ref{eq:cylinder_eq} may be used to estimate the magnitude of the generated anisotropy
for all radii, and Assumption 2 that the anisotropy vector components in the final state may be parametrized
using the initial state's linear density.

Given Assumptions 1 and 2, the anisotropy model consists of the equations,
\begin{subequations}\label{eq:anisotropy_model}
  \begin{eqnarray}
    \alpha &\equiv& 1 - \Big(\frac{I_2}{I_1}\Big)\Big(\frac{r_2}{r_1}\Big)\\
    \alpha_{rz}(r) &=& \Big(\frac{n_\beta(r)}{n}\Big)\alpha,\\
    \alpha_{r\theta}(r) &=& \Big(\frac{n_c(r)}{n}\Big)\alpha
\end{eqnarray}
\end{subequations}
where $n_\beta/n$ and $n_c/n$ are the radially varying betatron and cyclotron orbital density fractions
(\textit{cf.}~the prequel for the analytic fractions computed for the Bennett pinch~\cite{kinetics_i}).
The anisotropies appearing in Eqs.~\ref{eq:anisotropy_model} are defined as
\begin{subequations}\label{eq:anisotropy_definitions}
\begin{eqnarray}
  \alpha_{rz} &\equiv& 1 - \frac{T_z}{T_r},\label{eq:agyrotropic_anisotropy}\\
  \alpha_{r\theta} &\equiv& 1 - \frac{T_r}{T_\theta},\label{eq:gyrotropic_anisotropy}
\end{eqnarray}
\end{subequations}
and are referred to as the agyrotropic and gyrotropic anisotropies in the following.

\begin{figure}[h!]
  \includegraphics[width=0.5\linewidth]{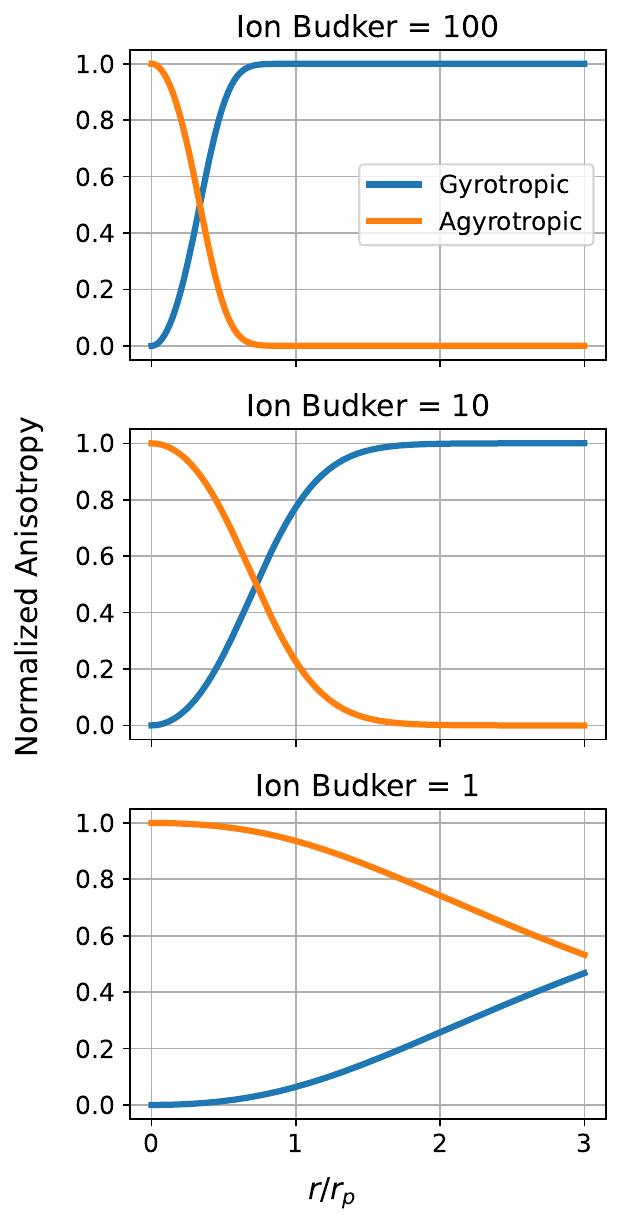}
  \caption{Hybrid CGL model (Eqs.~\ref{eq:gyrotropic_anisotropy} and~\ref{eq:agyrotropic_anisotropy})
    for ion anisotropy produced by collisionless flux compression for various ion Budker parameters.
    Gyrotropic anisotropy is induced by magnetic moment conservation for well-magnetized ions,
    while agyrotropic anisotropy is induced by collisionless betatron heating.
    The principal anisotropy axis rotates from field-aligned to current-aligned through the transitional magnetization
    layer, whose location and thickness is a function of the ion Budker parameter.
    \label{fig:anisotropy}}
\end{figure}

Figure~\ref{fig:anisotropy} shows the gyrotropic and agyrotropic ion anisotropy profiles predicted by
Eqs.~\ref{eq:anisotropy_model} for various ion Budker parameters.
The model predicts that agyrotropic anisotropy (induced by collisionless compression of ion betatron orbits)
is dominant for small Budker parameters,
while gyrotropic anisotropy (from collisionless compression of ion cyclotron trajectories) is dominant in the opposite limit.
In the intermediate regime, $\mathfrak{B}_i=\mathcal{O}(10)$, ion agyrotropy terminates
in the vicinity of the pinch radius, which has important consequences for the
transport of axial momentum and diamagnetic response of the compressing plasma discussed in the remainder of this work.

\subsection{Additional sources of agyrotropic anisotropy in the Z pinch}
In addition to adiabatic compression,
agyrotropic anisotropy may be induced non-adiabatically due to
collisionless shock heating~\cite{gedalin_collisionless_shock}
during fast Z pinch compression. 
Agyrotropy is induced in collisionless shock heating due to electric field gradients on the scale of the ion inertial length.
In addition, radial electric field gradients induced by entropy mode fluctuations~\cite{ricci_prl_2006}
and ensuing kinetic turbulence could also contribute to agyrotropy in ion distributions.
Agyrotropic heating is frequently observed in simulations of kinetic plasma turbulence
on ion inertial length scales~\cite{servidio_2015, Yang_2023}.
In this connection, note that species inertial length $\delta_s=c/\omega_{ps}$ is directly proportional
to pinch radius through the species Budker parameter~\cite{mahajan1989exact}; $\mathfrak{B}_s=\frac{r_p^2}{4\lambda_s^2}$.
Finally, strong electric fields can significantly modify particle orbits and drifts,
particularly around magnetic O-points~\cite{ilon_2021}.

\subsection{A particular example of betatron heating}
\begin{figure}[h!]
  \includegraphics[width=0.75\linewidth]{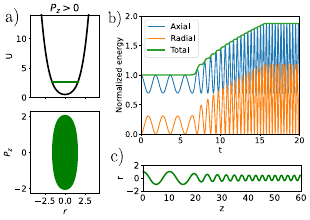}
  \caption{
    Demonstration of near-axis ion betatron heating from collisionless flux compression
    for a typical betatron orbit with $P_z>0$,
    with a) the effective potential $U(r)$ and radial action $J_r$
    (in green, the area enclosed by the orbit in $(r, P_r)$ phase-space),
    b) radial and axial components of kinetic energy during flux compression,
    and c) a segment of the accelerating betatron trajectory.
    The work done on the ion by the inductive electric field $E_z=-\partial_tA_z$ goes partly into
    the ion's axial drift energy and partly into its bounce (thermal) energies.
    Thus, the drift frame's axial bounce energy $\langle (v_z - \langle v_z\rangle)^2\rangle$
    heats less than the radial bounce and axial drift energies.
    \label{fig:betatron_heating1}}
\end{figure}

Figure~\ref{fig:betatron_heating1} presents a visual aid to the discussion of Section~\ref{sec:cgl_betatron},
presenting an example of betatron heating and acceleration in a compressing azimuthal magnetic flux.
An ion with $P_z>0$ is evolved in the model time-dependent $\hat{z}$-directed vector potential
$A_z = -A_0(t)\frac{r^2}{2a^2}$, with $a=1$ and $A_0(t)$ a specified flux compression function.
The ion is released from $r=1$ with $v_r=0$ and $v_z=1$, and the flux factor $A_0(t)$
is linearly and slowly varied from an initial value to a final value.

The inductive electric field $E_z=-\partial_tA_z$ does work on the ion, increasing
both the radial bounce energy and the axial drift/bounce energy in step (minus $P_z^2/2m$).
Most of the axial oscillation energy is channeled into the axial drift motion,
increasing the axial bounce energy in the drifting frame much less than the radial bounce energy,
because most of the axial energy in Fig.~\ref{fig:betatron_heating1} is composed of the
Galillean energy boost $\langle v_z\rangle \cdot \delta v_z$ from the oscillation-center frame to the lab frame.
Thus, betatron heating displays radial/axial temperature anisotropy because the change in radial bounce energy
is distinct from the change in axial bounce energy in the drifting frame.

During collisionless adiabatic flux compression of a large Larmor radius Z pinch,
mean ion betatron drift changes self-consistently with plasma current and density,
and mean radial bounce energy varies so that radial pressure balances the
self-magnetic energy $\frac{\mu_0}{8\pi}I^2$ (Bennett relation).
Axial temperature is unconstrained by radial force balance, but is constrained by kinetic equilibrium
considerations (which was not appreciated in early Z-pinch studies, see for example
the claim in Bennett 1934 that axial temperature is irrelevant to equilibrium~\cite{bennett_1934}).

\section{\label{sec:model-distributions}Kinetic equilibrium of magnetically self-focused flow}
Collisionless kinetic equilibrium is a state where the distribution of particle velocities
remains constant in time in the absence of collisions, unlike the fluid picture where collisions
are assumed to be frequent enough to maintain a near-Maxwellian velocity distribution.
Kinetic equilibria are stationary, self-consistent solutions to coupled kinetic-field equations,
\textit{e.g.}, the Vlasov-Maxwell system.
Collisionless plasmas find preferred equilibria through phase-mixing processes,
meaning the dynamical phase-volume preserving rearragement of phase fluid into a
lower energy state,~\textit{i.e.}~waves and instabilities\cite{Helander_Mackenbach_2024}.
Most kinetic equilibria are not such minimum-energy states.
Collisional kinetic equilibria, when relaxation and nonthermal forcing are in balance,
are the starting point of all transport closures, \textit{i.e.}, the zero-order state of the plasma is non-Maxwellian.

Collisionless kinetic equilibrium is any function of the constants of motion (Jeans' theorem).
As discussed in the prequel~\cite{kinetics_i}, for the flow pinch
these constants are the energy $H$ and axial canonical momentum $P_z$.
The distribution is assumed to be canonical in the energy ($e^{-\beta H}$) due to phase mixing and quasineutrality.
Recent work has shown that the maximum entropy state of the canonical ensemble is described by
the family of kappa distributions~\cite{kappa_distributions} which, although a universal description~\cite{Livadiotis_2024},
relies on a free parameter determined through a data fit.
This would unnecessarily complicate the present analysis, which is taken to its limit in the Boltzmann distribution.

The primary nonthermal features of the flow Z pinch arise in the distribution's dependence on canonical momentum.
In this case, an approach which we call Suzuki's Hermite method~\cite{suzuki2008novel} allows the construction of
self-consistent kinetic equilibria using the Hermite polynomials associated with the canonical energy
distribution.
Suzuki's Hermite method determines a generating function in the canonical momentum whose derivatives generate the fluid moments
self-consistently satisfying the field equations, ensuring a physically valid equilibrium.
An early study of the method is found in Channell~\cite{channell_1976}, and a recent thorough study appears in
Allanson et al~\cite{allanson2016one}.

We demonstrate the versatility of Suzuki's Hermite method by presenting several examples of
kinetic equilibria, including the well-known Bennett pinch equilibrium and a sheared-flow equilibrium
with anisotropic pressure. The latter is particularly interesting as it provides a framework for
understanding how velocity shear and temperature anisotropy interact in the collisionless regime.
In Section~\ref{sec:flows_sec} we describe this interaction in the context of collisionless gyroviscosity,
which could play a crucial role in the stability and dynamics of sheared flow Z pinches.
Here, we also explore the relationship between the generating function of Suzuki's method
and the plasma flow profile, revealing a deep connection between the kinetic and fluid descriptions of the Z pinch.




\subsection{\label{sec:vlasov-maxwell-solutions}Suzuki's Hermite method of kinetic equilibrium}
To begin, we present the aforementioned method~\cite{suzuki2008novel} to generate a self-consistent kinetic
equilibrium using the properties of Hermite polynomials.
Consider the ansatz distribution
\begin{equation}\label{eq:suzuki2008}
  f(A_z, H) = n_0\Big(\frac{\beta}{2\pi}\Big)^{3/2}
  \Big(\sum_{n=0}^\infty g_n(A_z)H_n(v_z)\Big)e^{-\beta H}
\end{equation}
with $H=\frac{m}{2}(v_x^2 + v_y^2 + v_z^2) + q\varphi$ the energy,
$\beta=(k_BT)^{-1}$ the inverse temperature,
$H_n$ the Hermite polynomials, $A_z$ the $z$-component of magnetic potential,
and $\varphi$ the electric potential.
Equation~\ref{eq:suzuki2008} is a kinetic equilibrium with $f=f(P_z,H)$ provided that
\begin{equation}\label{eq:substituted_equilibrium}
  \sum_{n=0}^\infty (g_n'(a_z)H_n(v_z) - g_n(a_z)H_n'(v_z)) = 0
\end{equation}
as found by substitution of Eq.~\ref{eq:suzuki2008} into the kinetic equation.
Here the vector potential is written in lower case when it takes on the units of velocity,
$a_z \equiv qA_z/m$, and it is normalized to the thermal speed $v_t$.
Equation~\ref{eq:substituted_equilibrium} is satisfied if each term of the series balances such that
\begin{equation}\label{eq:substituted_equilibrium2}
  \frac{dg_n}{da_z}H_n = g_n\frac{dH_n}{dv_z}.
\end{equation}
Now, the Hermite polynomials are the orthogonal family with the property $H_n'(x) = 2nH_{n-1}(x)$.
Substitution of the Hermite property into Eq.~\ref{eq:substituted_equilibrium2} shows that
the coefficient functions $g_n$ must satisfy the recurrence relation
\begin{equation}\label{eq:suzuki_relation}
  g_{n+1} = \frac{1}{2(n+1)}\frac{dg_n}{da_z}.
\end{equation}
Iteratively substituting Eq.~\ref{eq:suzuki_relation} into Eq.~\ref{eq:suzuki2008} gives the series
\begin{equation}\label{eq:suzuki_series}
  f = n_0\Big(\frac{\beta}{2\pi}\Big)^{3/2}
  \Big(\sum_{n=0}^\infty \frac{1}{2^n n!}\frac{d^ng_0}{da_z^n}H_n(v_z)\Big)e^{-\beta H}
\end{equation}
with $g_0=g_0(a_z)$ the first Hermite coefficient.
The particle density and current density follow as
\begin{subequations}
\begin{eqnarray}
n_\alpha(\varphi, a_z) &=& n_0 g_{0\alpha}(a_z)e^{-\beta_\alpha q_\alpha \varphi}, \label{eq:ng0}
\\
j_{z,\alpha} &=& q_\alpha n_0v_{t\alpha}\frac{dg_{0,\alpha}}{da_z}e^{-\beta_\alpha q_\alpha \varphi} \label{eq:jg1}
\end{eqnarray}
\end{subequations}
because $H_{0,1}(z) = 1, 2z$ are orthogonal to all other $H_n$ over the Gaussian weighting ($e^{-\beta H}$).
Equation~\ref{eq:suzuki_series} solves the kinetic equation, and its moments
consistently solve the Vlasov-Maxwell system provided it satisfies the nonlinear Poisson equations
\begin{subequations}
\begin{eqnarray}
  \nabla^2 \varphi &=& \varepsilon_0^{-1}\sum_\alpha q_\alpha n_{0\alpha}g_{0\alpha}(a_z)
                     e^{-\beta_\alpha q_\alpha\varphi},\label{eq:sp1}\\
  \nabla^2 A_z &=& \mu_0\sum_\alpha q_\alpha n_{0\alpha} v_{t\alpha}\frac{dg_{0\alpha}}{da_z}
                 e^{-\beta_\alpha q_\alpha\varphi}.\label{eq:sp2}
\end{eqnarray}
\end{subequations}
Thus, the function $g_0=g_0(a_z)$ is a generating function for the equilibrium,
determining density, current density, and distribution function
in the laboratory frame through its derivatives.
Recalling that axial magnetic vector potential is magnetic flux per unit length
($\psi_\theta' \equiv \int_0^r B_\theta(r')dr' = -A_z$), this means that the generating function
and all of its derivatives (the fluid moments) are flux functions.
Notably, the toroidal flux function in kinetic tokamak equilibrium can be treated analogously through an
azimuthal momentum ($P_\theta$) expansion~\cite{kaltsas_2024}.
Intuitively, the generating function of a distribution canonical in the energy
is expressed in Hermite series due to the properties of the Weierstrass transform,
for which $e^{-\beta H}$ is the kernel~\cite{allanson2016one}.
We now explore a few paradigmatic equilibria in terms of simple generating functions.




\subsection{Semi-canonical (shifted Maxwellian) equilibrium\label{subsec:semi_canon}}
The kinetic equilibrium of radially uniform axial velocity is the isothermal Bennett pinch,
whose distribution is semi-canonical (namely, a shifted Maxwellian).
The function $g_0(a_z)$ is exponential in $a_z$,
\begin{equation}\label{eq:max_shift}
  g_{0s}(a_z) = \exp(2m_s\beta_su_{0s}a_z).
\end{equation}
Repeated differentiation of Eq.~\ref{eq:max_shift} and application
of the Hermite generating function,
\begin{subequations}
\begin{eqnarray}
\frac{d^ng_{0s}}{da_z^n} &=& \Big(\frac{2u_0}{v_{ts}}\Big)^ng_0(a_z), \label{eq:gen1}
\\
\sum_{n=0}^\infty H_n(v_z)\frac{u_0^n}{n!} &=& \exp(2u_0v_z - u_0^2), \label{eq:gen2}
\end{eqnarray}
\end{subequations}
arrives at the shifted Maxwellian distribution, 
\begin{equation}
  f_s = n_{0s}\Big(\frac{\beta_s}{2\pi}\Big)^{3/2}e^{-\beta_s q_s(\varphi - 2u_{0s}A_z)}e^{-\beta_s m(v_x^2 + v_y^2 + (v_z - u_0)^2)/2}.
\end{equation}
Both the Bennett pinch (cylindrical) and Harris sheet (Cartesian) are generated by
Eq.~\ref{eq:max_shift} as they satisfy
(under quasineutrality in the center-of-charge frame) the Liouville equation
\begin{equation}\label{eq:liouville_pde}
  \nabla^2A_z = -Ke^{2A_z},
\end{equation}
with $K$ the curvature constant~\cite{conformal_equilibria}.
The cylindrical solution to Eq.~\ref{eq:liouville_pde} is the Bennett pinch's vector potential, given by
\begin{equation}\label{eq:bennett_potential}
  A_z = -A_0\ln(1 + (r/r_p)^2)
\end{equation}
where $r_p$ is the characteristic pinch radius and $A_0=\frac{\mu_0}{4\pi}I_\infty$ is the self-magnetic flux
per unit length ($I_\infty$ is the total current enclosed to infinity).
Regarding the generating function, it is useful that the following combination is unity,
\begin{equation}\label{eq:constants_unity}
  2q_s\beta_su_0A_0 = \frac{\mu_0}{4\pi}\frac{N_s q_s^2}{m_s}\Big(\frac{u_0}{v_{ts}}\Big)^2 = 1
\end{equation}
for each species $s$, where $N_s$ is the linear density (particles per unit length) of species $s$.
The first factor is the species Budker parameter $\mathfrak{B}_s$.
For the Bennett pinch, the inverse Budker parameter is the drift Mach number for that species~\cite{kinetics_i},
\begin{subequations}
\begin{eqnarray}
\mathfrak{B}_s &\equiv& \frac{q_s^2}{4\pi\epsilon_0}\frac{N_s}{m_sc^2}, \label{eq:budker}
\\
\mathfrak{B}_s^{-1} &=& \Big(\frac{u_0}{v_{ts}}\Big)^2. \label{eq:budker_identity}
\end{eqnarray}
\end{subequations}
From Eq.~\ref{eq:constants_unity}, it follows from the generating function that the Bennett pinch particle
and current densities are given by~\cite{meier_two_fluid}
\begin{subequations}
\begin{eqnarray}
n_s(r) &=& \frac{N_s}{\pi r_p^2}(1 + (r/r_p)^2)^{-2}, \label{eq:bennett_density}
\\
j_z(r) &=& \frac{I_\infty}{\pi r_p^2}(1 + (r/r_p)^2)^{-2}. \label{eq:bennett_current}
\end{eqnarray}
\end{subequations}
In the isothermal Bennett kinetic equilibrium, both ions and electrons have uniform flow velocities,
so their relative velocity is uniform too.
As long as the relative velocity of two species is radially uniform then the magnetic potential
will satisfy Eq.~\ref{eq:liouville_pde} and will be of the Bennett form, Eq.~\ref{eq:bennett_potential}.
The equilibrium has the important property that there is no radial electric field in the
center-of-charge frame~\cite{gratreau_orbits} where $u_i = -u_e$ (assuming $Z_i=1$).

\subsection{Sheared-flow, anisotropic (shifted bi-Maxwellian) equilibrium}
The generator of the next simplest equilibrium is quadratic in the flux,
namely $g_0 = \exp(c_0a_z + c_1a_z^2)$ for constants $c_0$, $c_1$.
This equilibrium consists of a two-temperature bi-Maxwellian distribution with an
agyrotropic pressure anisotropy supporting a radially sheared axial flow~\cite{del_sarto_aniso}.
Anisotropic collisionless equilibria with sheared flows were explored by Mahajan and Hazeltine~\cite{mahajan_shear},
whose discussion we summarize here in the context of the Z pinch.

This sheared anisotropic equilibrium is the simplest kinetic equilibrium displaying the
relationship between shear and anisotropy,
enabling an understanding of how these two phenomena interact in the collisionless regime,
which is relevant to Z pinches approaching fusion conditions.
All sheared-flow collisionless kinetic equilibria are ``inviscid'' because they represent a stationary flow profile,
which is a notable property.
However, as will be discussed later, the collisionless magnetized dynamics should still be considered highly viscous
in the sense that initially out-of-equilibrium flow profiles will quickly \textit{find} a kinetic equilibrium supporting a steady flow.

Proceeding by direct calculation from an ansatz distribution is the simplest way to show existence of the equilibrium
and determine its most fundamental properties.
Suppose the distribution function is a two-temperature (bi-)Maxwellian whose first velocity moment 
$\langle v_z\rangle = u_z(r)$ is radially sheared (with species subscripts suppressed),
\begin{equation}\label{eq:bi-maxwellian}
  f(r,\bm{v}) = \frac{\beta_\perp}{2\pi}\sqrt{\frac{\beta_z}{2\pi}}n(r)e^{-\beta_\perp \frac{m}{2}(v_r^2 + v_\theta^2)}e^{-\beta_z\frac{m}{2}(v_z - u_z(r))^2}.
\end{equation}
Substituting Eq.~\ref{eq:bi-maxwellian} into the Vlasov equation leads to
\begin{equation}\label{eq:bimax_vlasov}
  \partial_t f + v_r\Big(\frac{n'}{n} + \beta_\perp q\varphi' + m\beta_\perp u_z(r)\omega_c\Big)f + m\beta_z v_r(v_z-u_z)A_{rz}f = 0
\end{equation}
where $\omega_c=qB/m$, a shear stress drive term $A_{rz} \equiv u_z'-\alpha_{rz}\omega_c$, and
the distribution anisotropy is $\alpha_{rz} \equiv 1 - T_z/T_r$ (for $T_r=(k\beta_\perp)^{-1}$).
Taking the first radial moment of Eq.~\ref{eq:bimax_vlasov} yields
\begin{equation}
  \partial_t(n\langle v_r\rangle) = -n'kT_r -qn\varphi' - qnu_zB_\theta
\end{equation}
giving the equilibrium radial force balance $\nabla p = qn(\bm{E} + \bm{v}\times\bm{B})$
because $\langle v_r\rangle=0$.
The distribution function supports this force balance in equilibrium
(that is, with $\partial_tf=0$) if the driver $A_{rz}=0$,
namely when the anisotropy and shear profile are related by
\begin{equation}\label{eq:shear_requirement}
  \frac{du_z}{dr} = \alpha_{rz}\omega_c.
\end{equation}
Having assumed a radially constant anisotropy $\alpha_{rz}$,
the axial species velocity is then given by $u_z = u_0 - \alpha_{rz}a_z$,
meaning that the equilibrium velocity profile is a linear flux function.

The bi-Maxwellian distribution is arrived at from the generating function 
using the completeness relation for Hermite polynomials~\cite{wiener1988fourier},
as the derivatives of the generating function are themselves Hermite polynomials.
Applying the completeness relation
for the sum $\sum_n \frac{1}{n!4^n}H_n(u_0+\alpha_{rz}a_z)H_n(v_z)$ results in a bi-Maxwellian.
The algebra is simpler to proceed as above, and as conducted in Mahajan and Hazeltine~\cite{mahajan_shear},
who ultimately find the bi-Maxwellian equilibrium for the sheared Bennett equilibrium (following section) to be
\begin{equation}\label{eq:sheared_bennett_equilibrium}
  f(P_z,H) = \exp(-\beta(H + u_0P_z + \gamma P_z^2/2m))
\end{equation}
where $\gamma$ is a constant related to the anisotropy.
Equation~\ref{eq:sheared_bennett_equilibrium} is a nice closed form result for the distribution
function whose sheared flow is a linear flux function.
Higher-order dependence of the velocity on the flux function is explored in Section~\ref{sec:nonequilibrium},
but the series appear not to sum cleanly into exponentials as in Eq.~\ref{eq:sheared_bennett_equilibrium}.

\subsubsection{Sheared flow, anisotropic Bennett equilibrium}
The self-consistent two-species equilibrium has been presented in Channon~\cite{channon2001z}.
For reference, this solution is repeated here, 
and Channon's integration constants considerably simplified.  
In general, sheared two-species flows lead to magnetic potentials of a more complex form than
the Bennett profile's (Eq.~\ref{eq:bennett_potential}).
Nevertheless, if the electron and ion axial shear flows are equal then
their relative velocity remains radially uniform.
This is consistent with the bi-Maxwellian sheared equilibrium provided that the electron and ion agyrotropic anisotropies balance
as $\alpha_i/m_i + \alpha_e/m_e=0$.
For $\alpha_i>0$, the electrons are provided a comparatively gentle agyrotropy with $T_r < T_z$.
This balance of anisotropies together with quasineutrality leads to the Liouville equation for potential (Eq.~\ref{eq:liouville_pde})
and the magnetic potential retains the Bennett form (Eq.~\ref{eq:bennett_potential}).

The species densities, velocities, magnetic field, and radial electric field are given by
\begin{subequations}\label{eq:sheared_bennett}
\begin{eqnarray}
  n_s(r) &=& \frac{N_s}{\pi r_p^2}\frac{1}{(1 + (r/r_p)^2)^{2}}, \label{eq:bennett_density2}\\
  B_\theta(r) &=& \frac{\mu_0I_\infty}{2\pi r_p}\frac{r/r_p}{1 + (r/r_p)^2}, \label{eq:bennett_current2}\\
  \bm{v}_i(r) &=& u_0(1 + 2\alpha_i\mathfrak{B}_i\ln(1 + (r/r_p)^2))\hat{z}, \label{eq:ion_velocity}\\
  \bm{v}_e(r) &=& -u_0(1 - 2\alpha_i\mathfrak{B}_i\ln(1 + (r/r_p)^2))\hat{z}, \label{eq:electron_velocity}\\
  E_r(r) &=& -2\alpha_i\mathfrak{B}_i\ln(1 + (r/r_p)^2)u_0B_\theta(r),
\end{eqnarray}
\end{subequations}
where $\mathfrak{B}_i$ is the ion Budker parameter and $\alpha_i$ the ion agyrotropic anisotropy.
As a generalized Bennett equilibrium, all the usual relations apply, provided that the radial temperature $T_\perp$
is used to define the thermal velocity in the Bennett relation and Eq.~\ref{eq:budker_identity}.

\subsection{Higher-order kinetic equilibria of magnetically self-focused flows\label{sec:nonequilibrium}}
Collisionless kinetic equilibrium of a sheared-flow self-pinched plasma is not necessarily a bi-Maxwellian distribution,
which occurs for an isothermal pinch of radially constant $\alpha_{rz}$ temperature anisotropy in which flow velocity
is a linear function of magnetic flux.
Experimental and numerical simulation dynamics likely display both gyrotropic and agyrotropic anisotropies,
due both to flow shear and to the asymmetric responses of cyclotron and betatron trajectories to changes in flux;
hence general kinetic equilibria should display both axial and azimuthal temperature gradients.
Non-isothermal kinetic equilibria are reserved to a future work.

But even considering the isothermal solutions, velocity need not be a linear flux function.
Equilibria can be found as any polynomial function of flux.
This section studies more general kinetic equilibria such that the axial velocity
is some polynomial of the flux function, and studies the characteristic features of these solutions.
First observe that the axial velocity is given by a logarithmic derivative of
the generating function,
\begin{equation}\label{eq:velocity_g_relation}
  \langle v_z\rangle = \frac{1}{g_0}\frac{dg_0}{da_z} = \frac{d\ln g_0}{da_z}.
\end{equation}
Thus, each exponential polynomial $g_0(a_z)$
is associated with a velocity profile as a polynomial in the magnetic potential,
\begin{equation}\label{eq:power_generator}
  g_0(a_z) = \exp\Big(\sum_{n=0}^\infty \frac{c_n}{n+1}a_z^{n+1}\Big) \implies \langle v_z\rangle = \sum_{n=0}^\infty c_na_z^n
\end{equation}
where $c_n$ are some coefficients.
Equation~\ref{eq:power_generator} is a formal power series expansion of axial velocity about the pinch axis.  
With the velocity expressed as such a flux function (Eq.~\ref{eq:power_generator}),
we then seek the Hermite series coefficients to determine the equilibrium distribution function.

Substitution of Eq.~\ref{eq:power_generator} into Eq.~\ref{eq:suzuki_series}
produces the distribution $f=\widetilde{f}(P_z)e^{-\beta H}$ by the associated function of $P_z$,
whose Hermite series coefficients are the derivatives of $g_0(a_z)$.
Let
\begin{equation}\label{eq:h_def}
  h(a_z)\equiv \sum_{n=0}^\infty\frac{c_n}{n+1}a_z^{n+1}
\end{equation}
so that $g_0=\exp(h(a_z))$.
Application of Fa\`{a} di Bruno's formula to Eq.~\ref{eq:power_generator} yields the Hermite coefficients as
\begin{equation}\label{eq:bell_polynomial}
  \frac{d^ng_0}{da_z^n} = e^{h(a_z)}B_n(h',\cdots,h^{(n)})
\end{equation}
where $B_n$ is the n'th complete Bell polynomial~\cite{bell_polynomial}.
Thus, the semi-canonical distribution function associated with the postulated velocity profile of Eq.~\ref{eq:power_generator} is
\begin{equation}\label{eq:general_hermite_distribution}
  f = e^{-\beta H + h(a_z)}\sum_{n=0}^\infty\frac{1}{2^nn!}B_n(h',\cdots,h^{(n)})H_n(v_z).
\end{equation}

\subsubsection{Solution in multiple-variable Hermite polynomials}
The coefficients of the Hermite series in Eq.~\ref{eq:general_hermite_distribution} can be
expressed recursively using the multiple-variable Hermite polynomials originally introduced by Charles Hermite
and reintroduced by Dattoli et al~\cite{DATTOLI199471}.
We illustrate the general theory by considering a series for the axial velocity up to second order in the magnetic flux,
\begin{equation}\label{eq:cubic_order_in_velocity}
  v_z = u_0 - c_1a_z - a_z^2/2
\end{equation}
which will involve the three-variable Hermite polynomials.
Repeated differentiation of the generating function $g_0 = e^{h(a_z)}$ (Eq.~\ref{eq:h_def})
applied to Eq.~\ref{eq:cubic_order_in_velocity} yields the formula
\begin{equation}\label{eq:derivatives3}
  \frac{d^ng_0}{da_z^n} = e^{h(a_z)}\text{He}_n(u_0 - c_1 a_z - a_z^2, -c_1 - 2a_z, -1)
\end{equation}
where $\text{He}_n(x,y,z)$ is the three-variable Hermite polynomial, generated by the recurrence relation~\cite{DATTOLI199471}
\begin{equation}\label{eq:three_variable_hermite}
  \text{He}_n = x\text{He}_{n-1} + y(n-1)\text{He}_{n-2}
  + z(n-1)(n-2)\text{He}_{n-3}
\end{equation}
with initial condition $\text{He}_0=1$, yielding $\text{He}_1=x$, $\text{He}_2=x^2+y$, $\text{He}_3 = x^2 + 3xy + 2z$, etc.
The usual single-variable Hermite polynomials are recovered with $\text{He}_n(x,-1,\cdots,0) = \text{He}_n(x)$.
Thus, the Hermite series for the function of canonical momentum corresponding to Eq.~\ref{eq:cubic_order_in_velocity} is
\begin{equation}\label{eq:general_cubic_hermite}
  \widetilde{f}(P_z) = e^{h(a_z)}\sum_{n=0}^\infty\frac{H_n(v_z)}{2^nn!}\text{He}_n(u_0-c_1a_z-a_z^2, -c_1 -2a_z, -1).
\end{equation}
Higher-order vector potential dependence involves higher-order Hermite polynomials.
That is, velocity as a cubic flux function involves four-variable Hermite polynomials, etc.,
generalizing the pattern of Eq.~\ref{eq:three_variable_hermite} in an obvious way.

In Eq.~\ref{eq:general_cubic_hermite}, the coefficients of the Hermite series are related to the Airy polynomials $\text{Pi}_n(x)$
introduced by Torre for applications in beam optics~\cite{Torre_2012} (by $\text{Pi}_n(x) = H_n(x,0,-1/3)$).
Equation~\ref{eq:general_cubic_hermite} shows that nonthermality of the equilibrium
where velocity is a quadratic flux function is described by a cross-series of Airy-like and Hermite polynomials.
A closed form of this series is unknown to the authors.

\begin{figure}[h!]
  \includegraphics[width=0.5\linewidth]{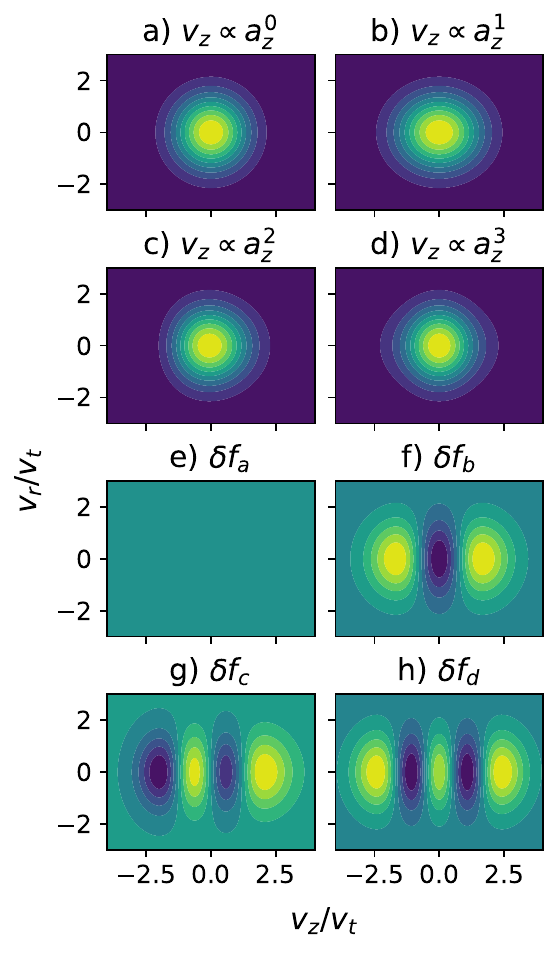}
  \caption{Contour plots of the ion distribution function $f_i$ in velocity space $(v_r, v_z)$ for (a-d) four different
    kinetic equilibria: (a) the semi-canonical (shifted Maxwellian) solution of uniform velocity,
    (b) the bi-Maxwellian equilibrium when $v_z$ is a linear flux function,
    (c) the skewed distribution for sheared flow with velocity a quadratic flux function, and
    (d) the kurtotic distribution for velocity as a cubic flux function.
    Parts (e-h) show the corresponding differences from the Maxwellian $\delta f = f - f_0$, highlighting that the higher-order
    velocity profiles are associated with finer velocity space gradients, encoding deviations of higher moments
    from equilibrium.
    The distributions are shown in the frame drifting at the mean velocity, and the color bars are normalized
    to the maximum values.
\label{fig:flux_functions}}
\end{figure}

The equilibrium was numerically investigated for quadratic and cubic velocity flux functions.
These were found to induce skewness and kurtosis, respectively, in the distribution function.
Figure~\ref{fig:flux_functions} shows numerically computed distributions (in their drifting frames)
computed for Fig.~\ref{fig:flux_functions}a) unsheared flow and Fig.~\ref{fig:flux_functions}b)-d) axial velocity
as a linear, quadratic, and cubic flux functions, respectively.
The coefficients were found using recursion relation for the multiple-variable Hermite polynomials,
similar to Eq.~\ref{eq:general_cubic_hermite}.

These kinetic equilibrium solutions
demonstrate that sheared flows are stationary in Z-pinch plasmas
only if the distribution function has certain non-thermal features.
If temperatures are isothermal and the distribution is canonical in the energy, though possibly anisotropic,
then the equilibrium velocity profile is a flux function.
First-order dependence on the flux is associated with pressure anisotropy (Fig.~\ref{fig:flux_functions}b),
quadratic dependence with skewness (Fig.~\ref{fig:flux_functions}c), and so on.
Regarding the heat flux, observe that a sum of Figs.~\ref{fig:flux_functions}b) and c) is
a close match to the equilibrium distribution shown in Fig.~7 of Malara 2022~\cite{Malara_2022}.
The interplay between velocity shear and pressure anisotropy has important consequences for
collisionless viscosity in the flow pinch. 


\section{\label{sec:flows_sec}Dynamical kinetic gyroviscosity: self-organization \& self-generation of magnetically self-focused sheared flow}
The previous section studied collisionless kinetic flow-pinch equilibria,
highlighting a sheared-flow, anisotropic-pressure equilibrium which paradigmatically illustrates
the interplay of sheared flows and pressure anisotropy.
The existence of a sheared-flow equilibrium is surprising from the perspective of unmagnetized media.
Collisionless unmagnetized flows, \textit{i.e.}~free molecular flows, are highly viscous;
stationary, finite-pressure sheared flows do not exist in the large Knudson number limit
because the flow-transverse components of trajectories are free to phase mix.
The magnetic field enables sheared equilibria through a counteracting gyrophase mixing~\cite{Gedalin_2015}
which tends to reduce agyrotropic anisotropy by smoothing out variations with respect to gyrophase angle.
These equilibrium solutions are valid regardless of orbit magnetization.

This section explores the collisionless limit through the interplay of flow shear and pressure anisotropy
which regulates non-equilibrium flows through the magnetized pressure-strain interaction.
This can be understood as the collisionless manifestation of gyroviscosity.
Analysis of the stress tensor reveals that magnetized flows respond to shear stress by radial transport of
axial momentum, as in the usual viscous process, but momentum transport is inhibited close to kinetic equilibrium.
Thus axial flows self-organize as a consequence of collisionless gyroviscosity,
which does not fully damp out flow shear but rather regulates it towards a stationary state,
\textit{i.e.}~a magnetized sheared-flow kinetic equilibrium.

Shear stress results from two kinds of forcing: from a driven flow, or from a driven anisotropy.
We term the former the forward process, in which viscosity mixes an initially thermal sheared flow
into a steady anisotropic sheared flow, freezing flow into a flux function.
The inverse process, namely generation of sheared flow, occurs from pressure anisotropy driven,
for example, through flux compression.
Self-generation of sheared flow by viscosity was, to our knowledge, first anticipated
by Velikhov~\cite{velikhov_1964} and Haines~\cite{haines_1965} in the context of the dynamic $\theta$-pinch.
Sheared-flow generation in the compressing Z pinch was studied with a hybrid Vlasov-fluid model
by Channon and colleagues~\cite{channon_1999, channon_2004}.
After discussing the forward process, we elaborate here on the flow-generating inverse process
discussed by Velikhov, Haines, and Channon et al.

Only from the point-of-view of kinetic equilibria are collisionless cross-field flows inviscid.
The kinetic dynamics may still be regarded as highly viscous in the sense
that a distribution supporting out-of-equilibrium flow rapidly mixes towards a ``frozen'' equilibrium,
thus organizing and regulating flow rather than fully dissipating it, but developing non-thermal features in the process.
The dynamical perspective clarifies the physics of the response of a current-carrying plasma to pressure fluctuations.

Dynamical collisionless gyroviscosity is a reactive dissipation in analogy to reactive resistance in electrical circuits.
In fact, dynamical gyroviscosity can induce velocity dispersion in one species and not another when ion and electron magnetizations
are decoupled through the mass ratio.
In this way, it will be seen that dynamical gyroviscosity produces reactive electrical resistance in response to
sufficiently rapid pressure fluctuations, which is associated with spatially varying pressure anisotropies.
The usual notion of dissipative gyroviscosity (\textit{i.e.}, non-reactive, or quasi-static) arises through collisional smoothing of
non-thermal features in the small Hall parameter limit (with respect to the dynamical scales, \textit{cf}. Section~\ref{sec:kinetic-regime}).


\subsection{Suppression of shear stress by pressure anisotropy in a magnetic field}
It is intuitive to first consider the forward process, namely viscous damping of sheared flows.
To begin, we extend the equilibrium analysis of Mahajan and Hazeltine
by considering the evolution of an agryotropic distribution initially out-of-equilibrium,
that is, when the equilibrium anisotropy condition $A_{rz}=0$ is not met.
Assuming initial radial force balance and introducing a collision operator $C[f]$, Eq.~\ref{eq:bimax_vlasov} can
be written as
\begin{equation}\label{eq:kinetic_equation}
  \frac{\partial f}{\partial t}\Big|_{t=t_0} = C[f] - m\beta_z v_r(v_z-u_d)A_{rz}f
\end{equation}
instantaneously at the initial time $t=t_0$.
Given the bi-Maxwellian form of $f$, the collisionless term on the RHS of Eq.~\ref{eq:kinetic_equation}
may be rewritten as derivatives of $f$, giving
\begin{equation}\label{eq:unthermal_operator}
  \frac{\partial f}{\partial t}\Big|_{t=t_0} = C[f] + \frac{\partial^2}{\partial v_r\partial v_z}(v_{th,r}^2A_{rz}f)
\end{equation}
and revealing it to be a diffusive operator, where $v_{th,r}^2 = kT_r/m$ is the radial thermal velocity.
The mixed-derivatives term of Eq.~\ref{eq:unthermal_operator}
satisfies many of the features of a collision operator, such as conservation of particles, energy, and momentum,
but drives shear stresses by not conserving the off-diagonal pressure tensor components $\Pi_{rz} = \int v_rv_zfd\bm{v}$.

\begin{figure}[h]
  \includegraphics[width=0.5\linewidth]{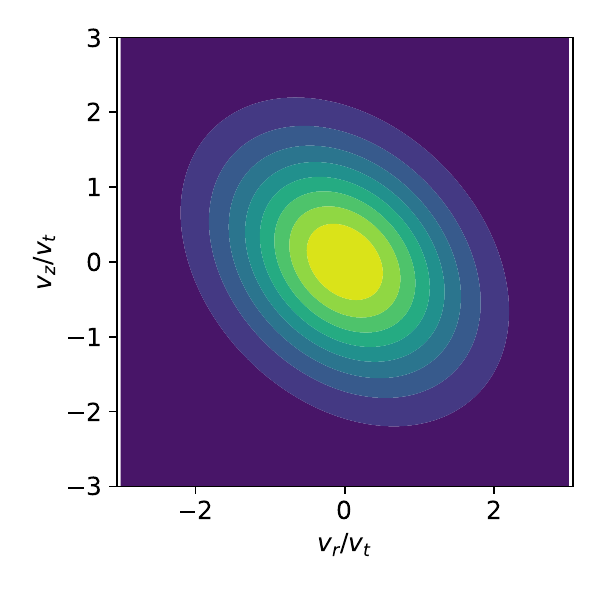}
  \caption{Exaggerated example of an anisotropic distribution function evolving
    from the off-diagonal diffusion in Eqs.~\ref{eq:diffusion_equation} (with $\nu=0$).
    The driving terms $A_{rz}=v_z' - \alpha_{rz}\omega_c$ produce shear stress,
    which appears as a sheared distortion of the distribution function in velocity space.
\label{fig:off_diagonal_diffusion}}
\end{figure}

Adopting a simple collision model such as the Lenard-Bernstein form shows 
the mixed derivatives to introduce off-diagonal terms to the velocity space diffusion matrix.
The Fokker-Planck equation then has the paradigmatic form
\begin{subequations}\label{eq:diffusion_equation}
\begin{eqnarray}
    \frac{df}{dt} &=& -\nu \nabla_v\cdot(\bm{v}f) + v_t^2\nabla_v\cdot(\mathcal{D}\nabla_v f),\\
    \mathcal{D} &=& \begin{bmatrix} \nu & A_{rz}/2\\ A_{rz}/2 & \nu\end{bmatrix}
\end{eqnarray}
\end{subequations}
where $\nu$ is the relaxation rate and $A_{rz}$ the shear stress drive frequency.
The off-diagonal terms produce shearing in the velocity space
(demonstrated in Fig.~\ref{fig:off_diagonal_diffusion})
as they are symmetric.

Shear stress generation by $A_{rz}\neq 0$ is observed by taking the moment for $\Pi_{rz}$ of Eq.~\ref{eq:unthermal_operator},
\begin{equation}\label{eq:off_diagonal_stress}
  \frac{\partial\Pi_{rz}}{\partial t}\Big|_{t=t_0} = A_{rz}nkT_r.
\end{equation}
Assuming the distribution to relax on a timescale $\tau$ (or observing the dynamics a short time later),
the solution to Eq.~\ref{eq:off_diagonal_stress} can be estimated as $\Pi_{rz} = (v_z' -\widetilde{\alpha}\omega_c)nkT_r \tau$
where $\widetilde{\alpha}$ is a relaxed anisotropy.
The radial viscous stress (from $n^{-1}\nabla\cdot\Pi$) is then 
\begin{equation}\label{eq:visc_stress}
  \frac{1}{nr}\frac{\partial r\Pi_{rz}}{\partial r} = \frac{1}{nr}\frac{\partial}{\partial r}\Big(nkT_r\tau r\Big(\frac{\partial v_z}{\partial r} - \widetilde{\alpha}\omega_c\Big)\Big).
\end{equation}
Ignoring the density gradient and assuming anisotropy is constant radially
produces the axial momentum transport equation,
\begin{equation}\label{eq:viscosity}
  \frac{\partial v_z}{\partial t} \sim \frac{1}{r}\frac{\partial}{\partial r}\Big(v_{tr}^2\tau
  r \frac{\partial }{\partial r}(v_z - \widetilde{\alpha}a_z)\Big)
\end{equation}
where $a_z=qA_z/m$ is the vector potential measured with units of velocity.
Equation~\ref{eq:viscosity} demonstrates that the condition $v_z' = \alpha_{rz}\omega_c$
is just the equilibrium state around which dynamic processes operate.

\subsection{Pressure-strain interaction and collisionless gyroviscosity}
We wish to understand how a distribution might ``find'' this collisionless
sheared-flow equilibrium from an initial non-equilibrium state.
This section shows that an initially out-of-equilibrium distribution does not find equilibrium,
but attains a decaying state with similar properties.
This occurs as follows: the out-of-equilibrium state generates shear stresses (Eq.~\ref{eq:off_diagonal_stress}).
These shear stresses produce pressure anisotropy and induce a viscous evolution of the plasma flow and current profiles.
Shear stress production is inhibited (saturates) if the state $A_{rz}=0$ is found.
As observed by Del Sarto, residual shear stress in this state means that radial transport of axial momentum
is mediated by magnetosonic wave emission~\cite{Sarto_2017}.

The discussion leading to Eq.~\ref{eq:viscosity} did not describe how pressure anisotropization
is induced by Eqs.~\ref{eq:diffusion_equation} because the assumed bi-Maxwellian distribution has zero shear stress.
A more general theory is found by direct analysis of the Vlasov equation without an ansatz distribution.
Del Sarto and Pegoraro performed such an analysis (neglecting heat fluxes) and
derived an elegant matrix equation for the collisionless evolution of the ion pressure tensor~\cite{del_sarto_how_anisotropization},
\begin{equation}\label{eq:full_pressure_tensor_equation}
  \frac{d\Pi}{dt} = [B + W, \Pi] - \{D,\Pi\} + 5\mathcal{C}\Pi.
\end{equation}
In Eq.~\ref{eq:full_pressure_tensor_equation}, $[\cdot,\cdot]$ represent matrix commutators and $\{\cdot,\cdot\}$ anti-commutators.
The matrix $\Pi$ is the pressure tensor, the matrices $W$ and $B$ are the vorticity and (signed) cyclotron frequency dual tensors
defined by $\omega_i = \varepsilon_{ijk}W_{jk}$ and $\omega_{c,i} = \varepsilon_{ijk}B_{jk}$, the matrix
$D = \frac{1}{2}(\nabla\bm{u} + (\nabla\bm{u})^T - 2/3(\nabla\cdot\bm{u})I)$ is the incompressible rate of shear tensor, and
$\mathcal{C} = -\frac{1}{3}(\nabla\cdot\bm{u})$ is the isotropic compression.
The matrix $D$ is the traceless symmetric part of the velocity gradient (strain rate) tensor.

We extend the discussion of Del Sarto and Pegoraro to analyze sheared flow about the axis of a
magnetically self-focused flow by a direct calculation.
Working in the $(r,z)$-plane with sheared axial flow and taking the azimuthal angle
to be an ignorable coordinate, the Cartesian theory is applicable provided that $T_\theta = T_r$
and the matrices are
\begin{subequations}\label{eq:matrices_shear}
\begin{eqnarray}
  D &=& \frac{v_z'}{2}\begin{bmatrix} 0 & 1 \\ 1 & 0\end{bmatrix},\\
  W &=& \frac{v_z'}{2}\begin{bmatrix} 0 & 1 \\ -1 & 0\end{bmatrix},\\
  B &=& \omega_c\begin{bmatrix} 0 & -1 \\ 1 & 0\end{bmatrix}.
\end{eqnarray}
\end{subequations}
Using Eqs.~\ref{eq:matrices_shear}, the commutators work out to
\begin{subequations}\label{eq:commutators}
\begin{eqnarray}
  [W,\Pi] &=& \frac{v_z'}{2}\begin{bmatrix} 2\Pi_{rz} & \Pi_{zz}-\Pi_{rr} \\ \Pi_{zz}-\Pi_{rr} & -2\Pi_{rz}\end{bmatrix},\\
  {[B,\Pi]} &=& \omega_c\begin{bmatrix} -2\Pi_{rz} & \alpha_{rz}\Pi_{rr} \\ \alpha_{rz}\Pi_{rr} & 2\Pi_{rz}\end{bmatrix},\\
  \{D,\Pi\} &=& \frac{v_z'}{2}\begin{bmatrix} 2\Pi_{rz} & \Pi_{zz}+\Pi_{rr} \\ \Pi_{zz}+\Pi_{rr} & -2\Pi_{rz}\end{bmatrix},
\end{eqnarray}
\end{subequations}
where again $\alpha_{rz}=1 - \Pi_{zz}/\Pi_{rr}$ is the agyrotropy coefficient defined previously,
which are substituted into Eq.~\ref{eq:full_pressure_tensor_equation}.

\subsubsection{Viscous dissipation of collisionless unmagnetized flow\label{sec:free_flow}}
For intuition, consider first the collisionless unmagnetized case with $\omega_c=0$ (essentially free molecular flow),
in which case
\begin{equation}\label{eq:unmagnetized_viscosity}
  \frac{d\Pi}{dt} = -v_z'\begin{bmatrix}0 & \Pi_{rr}\\ \Pi_{rr} & 2\Pi_{rz}\end{bmatrix}.
\end{equation}
The components of Eq.~\ref{eq:unmagnetized_viscosity} show that temperature anisotropy develops during
relaxation of collisionless sheared flow as the evolving shear stress $\frac{d\Pi_{rz}}{dt} = -v_z'nkT_r$ drives diffusion through
the viscous momentum flux $\frac{\partial v_z}{\partial t} = \frac{1}{n}\frac{\partial\Pi_{rz}}{\partial r}$,
and concurrently axial temperature increases as $\frac{dT_z}{dt} = -2v_z'\Pi_{rz}/n$.
The anisotropy occurs as, in the collisionless limit of viscosity, heating occurs only in the flow direction.
This axial heating is, of course, purely a consequence of the ballistic phase mixing associated with radial temperature.
The process continues until $v_z'\to 0$ and the shear stress asymptotically converges to a constant value.
Fetsch and Fisch have recently noted that such anisotropic heating in fact increases fusion reactivity~\cite{fetsch_2024}.

\subsubsection{Gyroviscous mixing of collisionless magnetized flow\label{sec:gyroviscous_damping}}
The magnetic field introduces gyrophase mixing, tending to reduce the gyrophase variations induced by the axial heating
due to mixing of nearby trajectories in Section~\ref{sec:free_flow}.
With $\omega_c\neq 0$, Eq.~\ref{eq:unmagnetized_viscosity} is instead,
\begin{equation}\label{eq:gyroviscosity_effect}
  \frac{d\Pi}{dt} =
  -\begin{bmatrix}
     2\omega_c\Pi_{rz} & (v_z' - \alpha_{rz}\omega_c)\Pi_{rr}\\
     (v_z'-\alpha_{rz}\omega_c)\Pi_{rr} & 2(v_z'-\omega_c)\Pi_{rz}
   \end{bmatrix}.
\end{equation}
Consider the collisionless evolution of an initially isotropic sheared-flow Z pinch with $\alpha_{rz}=0$.
Shear stress $\Pi_{rz}$ develops from the drive term $\frac{d\Pi_{rz}}{dt}=-v_z'\Pi_{rr}$, 
but also produces pressure anisotropy through the component equations
$\frac{d\Pi_{rr}}{dt}=-2\omega_c\Pi_{rz}$ and $\frac{d\Pi_{zz}}{dt}=-2(v_z'-\omega_c)\Pi_{rz}$,
which combine into,
\begin{equation}\label{eq:anisotropy_generation}
  \frac{d\alpha_{rz}}{dt} = -\frac{d}{dt}\Big(\frac{\Pi_{zz}}{\Pi_{rr}}\Big) =
  -4\frac{\Pi_{rz}}{\Pi_{rr}}((\omega_c - v_z'/2) - \alpha_{rz}\omega_c/2)
\end{equation}
Equation~\ref{eq:anisotropy_generation} shows that $\omega_c > v_z'/2$ is the critical condition for
the rate of viscous radial heating to exceed axial heating,
thereby producing anisotropy $\alpha_{rz} > 0$ and slowing the rate of shear stress production through
$\frac{d\Pi_{rz}}{dt}=-(v_z'-\alpha_{rz}\omega_c)\Pi_{rr}$.
Physically, this indicates that the ion gyrofrequency exceeds the angular velocity of fluid element rotation
in the sheared flow ($\omega_c > v_z'/2$).
Shear stress production is minimized if the state $v_z'=\alpha\omega_c$ is found,
but equilibrium will not be found (as heating continues in the radial and transverse directions)
without a dissipative process to remove the shear stress.
This appears to produce a dynamic oscillation which pumps energy into the radial pressure, 
leading to sustained magnetosonic wave emission~\cite{Sarto_2017}.




In other words, inspection of Eq.~\ref{eq:gyroviscosity_effect} suggests that the kinetic equilibrium
satisfying $v_z'=\alpha_{rz}\omega_c$ with $\Pi_{rz}=0$ (the bi-Maxwellian) is not an attractor 
as kinetic dynamics do not drive the plasma to a static state.
However, the condition $v_z'=\alpha_{rz}\omega_c$ should be dynamically attained as it minimizes
the time-evolving shear stresses through gyrophase mixing.
A characteristic signature of this reorganization should be radial magnetosonic wave emission.
Thus, collisionless gyroviscosity encourages fluid vorticity to be related to the gyrofrequency
through phase mixing.
In the isothermal limit, this makes the flow a flux function.
Self-consistent Vlasov-Maxwell simulations are planned to better understand this nonlinear process.

\subsection{Self-generated sheared axial flow: inverse gyroviscosity\label{sec:spon_sheared_flow}}
Modifying the thought experiment of Section~\ref{sec:gyroviscous_damping}
into an initially agyrotropic ($\alpha_{rz}\neq 0$) but quasi-static (shear-free, $v_z'=0$) plasma, we find that
the inverse process of viscosity occurs: a sheared axial flow is spontaneously generated.
Haines~\cite{Haines_2011} (above their Eq.~3.33) has commented on the possibility of
spontaneous sheared axial flow from finite Larmor orbit effects,
and Channon~\cite{channon_1999, channon_2004} has observed the effect in hybrid Vlasov-fluid simulations
of pinch compression.
In this section, we thoroughly discuss the phenomenon of sheared flow generation via pressure anisotropy,
as induced by adiabatic compression of betatron orbits in the case of Channon's results.

The shear stress is estimated from Eq.~\ref{eq:unthermal_operator}
after a short time $\tau$ as $\Pi_{rz}\approx \alpha_{rz}(\omega_c\tau)\Pi_{rr}$.
The axial component of the momentum equation now reads
\begin{equation}\label{eq:force_on_plasma}
  \frac{\partial (nv_z)}{\partial t} = -(\nabla\cdot\Pi)_z = -\frac{1}{r}\frac{\partial(r\alpha_{rz}(\omega_c\tau)\Pi_{rr})}{\partial r}.
\end{equation}
Equation~\ref{eq:force_on_plasma} can be understood as an axial force arising
by the gyroviscous stress of induced agyrotropic anisotropy.  
There is zero net force ($F_\text{net}=\int r^{-1}\partial_r(r\Pi_{rz})2\pi rdr = 0$) as the shear stress decays much faster than $r$,
thus conserving total momentum like any viscosity.

The axial force density (Eq.~\ref{eq:force_on_plasma}) is approximated
by taking the magnetic field and pressure to be Bennett distributed, and the 
agyrotropic anisotropy $\alpha_{rz}$ is modeled as due to betatron heating (Eq.~\ref{eq:agyrotropic_anisotropy})
and thus proportional to the density of betatron trajectories.
Under these assumptions, the shear stress after time $\tau$ is proportional to the betatron fluid pressure
and local cyclotron frequency,
\begin{equation}\label{eq:betatron_shear_stress}
  \Pi_{rz} \approx \alpha n_{\beta} kT \omega_{c}\tau
\end{equation}
with $\alpha$ the constant anisotropy magnitude (predicted by the CGL model),
$n_{\beta}(r)$ the local density of ion betatron trajectories, and $\omega_c(r)$ the local ion cyclotron frequency.
Equation~\ref{eq:betatron_shear_stress} should be compared to the collisional closure $\Pi_{rz}=-pv_z'\tau$
with $p$ the total scalar pressure~\cite{miller_2016}.

\begin{figure}[h!]
  \includegraphics[width=0.5\linewidth]{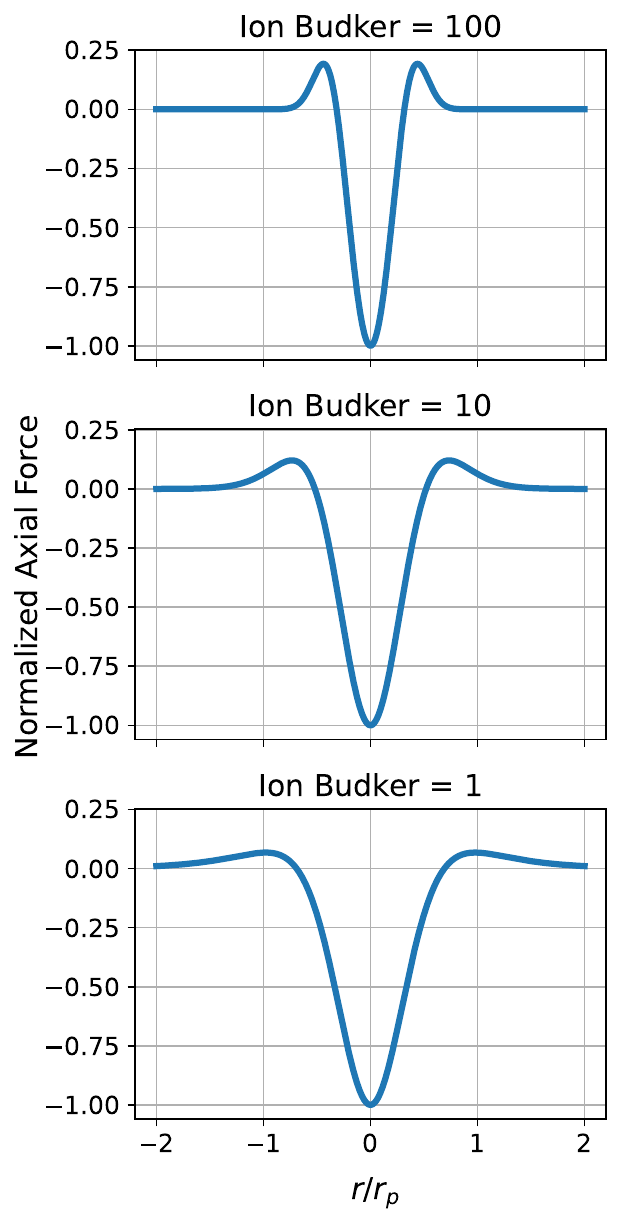}
  \caption{Normalized axial force on the ions in a Bennett pinch in response to (highly idealized) collisionless compression,
    for various ion Budker parameters.
    The force is induced by collisionless gyroviscosity arising from agyrotropic anisotropy,
    and is proportional to the pressure of ions following betatron trajectories.
    Spontaneous flow generation by anisotropy production can be considered an inverse gyroviscosity.
    The net force on the plasma is zero, hence a differential shear flow is generated that reverses direction
    at the edge of the transitional magnetization layer within a diffuse flow pinch.
\label{fig:axial_force_on_plasma}}
\end{figure}

Figure~\ref{fig:axial_force_on_plasma} plots the anticipated axial force
(from Eqs.~\ref{eq:force_on_plasma} and~\ref{eq:betatron_shear_stress}) for various ion Budker parameters (Eq.~\ref{eq:budker}),
showing the predicted oscillatory flow to reverse direction at the edge of the betatron density, or in other words
towards the limit of the transitional magnetization region.
This prediction agrees with the varying radial limit of sheared flow with large Larmor radius parameter (our Budker parameter)
observed by Channon~\cite{channon_2004}.
The generation of flow is nothing other than a viscous response to a pressure fluctuation, but is somewhat surprising at first
glance as the plasma is initially static.

In our situation of a quasineutral large ion Larmor radius current-carrying plasma, essentially only the betatron ions experience this
viscous response.
The electrons mostly follow cyclotron trajectories and heat gyrotropically.
It is interesting to observe that the force on the ions in response to the pressure change
is counter the electric current.
Thus, electrical current diffuses outwards.
Electric current diffusion leads to magnetic diffusion, and hence is resistive.
Thus, the self-generated ion flow resulting from flux compression illustrates an electrical resistance due purely to viscosity.
This resistance is reactive in the sense that it arises from dynamic changes in the plasma.
We suspect that the reactive gyroviscoresistivity described here may play a so-far underappreciated role as a dissipative process in dynamic Z pinches.

\subsection{Spontaneous rotation from an axial magnetic field}
Section~\ref{sec:spon_sheared_flow} observed that gyroviscosity 
drives spontaneous sheared axial flow due to agyrotropic anisotropy from \textit{e.g.}, collisionless betatron heating.
One naturally wonders if similar phenomena arise from the gyrotropic anisotropy imbued to the cyclotron fluid.
The answer is affirmative given a small frozen-in axial magnetic field~\cite{lucibello_zpinch},
in which case a spontaneous plasma rotation is predicted.
Similarly to the betatron heating case, such a response is reactively gyroviscoresistive with respect to the azimuthal/poloidal currents.
The differential rotation of a compressing $\theta$-pinch is somewhat well-known and was first predicted by Velikhov~\cite{velikhov_1964}.
In the general picture, doubly adiabatic compression of a screw pinch leads to differential rotation within its magnetized periphery
and differential axial flow within its current-dense core, as observed by Channon~\cite{channon_2004}.

Consider the pressure tensor dynamics with an axial magnetic field, assumed to be much smaller than the Z-pinch azimuthal flux.
To be cautious in cylindrical coordinates, we directly derive the equation
of Del Sarto and Pegoraro (Eq.~\ref{eq:full_pressure_tensor_equation}) from the polar Vlasov equation~\cite{vogman_polar},
taking a separable ansatz that $f=n(r)g(v_r,v_\theta-u_\theta(r),v_z-u_z(r))$ with $n(r)$ a quasineutral plasma density
and $u_\theta(r)$, $u_z(r)$ azimuthal and axial flow velocities.
The resulting kinetic equation is
\begin{eqnarray}\label{eq:polar_vlasov}
  \partial_tf + \frac{n'}{n}v_rf + \Big(v_\theta\omega_z - v_z\omega_\theta + \frac{v_\theta^2}{r}\Big)\partial_{v_r}f\nonumber\\
  -\Big(\frac{v_rv_\theta}{r} + v_r(u_\theta' + \omega_z)\Big)\partial_{v_\theta}f + v_r(\omega_\theta - u_z')\partial_{v_z}f = 0
\end{eqnarray}
with $\omega_z$ and $\omega_\theta$ the cyclotron frequencies due to the $z$- and $\theta$-oriented magnetic fields.
Taking moments for the pressure tensor ($\Pi_{ij} = \int v_iv_jf(v)d^3v$) and neglecting all third and fourth-order moments
(i.e., heat fluxes $Q_{ijk} = \int v_iv_jv_k f(v)d^3v$) does obtain Del Sarto and Pegoraro's result,
\begin{subequations}\label{eq:rotation_inducing}
  \begin{eqnarray}
    \partial_t\Pi_{rr} &=& 2(\omega_z\Pi_{r\theta} - \omega_\theta\Pi_{rz}),\\
    \partial_t\Pi_{\theta\theta} &=& -2(\omega_z+u_\theta')\Pi_{r\theta},\\
    \partial_t\Pi_{zz} &=& -2(\omega_\theta - u_z')\Pi_{rz},\\
    \partial_t\Pi_{rz} &=& \omega_z\Pi_{\theta z} - (u_z' - \alpha_{rz}\omega_\theta)\Pi_{rr},\\
    \partial_t\Pi_{r\theta} &=& -\omega_\theta\Pi_{\theta z} - (u_\theta' + \alpha_{r\theta}\omega_z)\Pi_{rr},\label{eq:rot1}\\
    \partial_t\Pi_{\theta z} &=& -(u_\theta' + \omega_z)\Pi_{rz} - (u_z' - \omega_\theta)\Pi_{r\theta},
  \end{eqnarray}
\end{subequations}
with $\alpha_{r\theta} = 1 - T_\theta/T_r$ the gyrotropic anisotropy, which we model with Eq.~\ref{eq:gyrotropic_anisotropy}
to occur typically far from the magnetic axis.
The cylindrical terms (such as Coriolis force) do appear to play a role in heat flux dynamics.
Given gyrotropic anisotropy, the axial magnetic field induces gyrotropic shear stress (Eq.~\ref{eq:rot1}),
inducing rotational force on the plasma in a manner similar to Eq.~\ref{eq:force_on_plasma}.
As a diffusion, the spontaneous rotation is differential, reversing direction radially.

\subsection{\label{subsec:char_anisotropy}Characteristic anisotropy of Alfv\'{e}nic sheared flow}
Sheared flows are thought to decrease the growth rate of kink instabilities of flow pinches~\cite{shumlak_shear},
particularly in combination with viscoresistive effects~\cite{G_O_Spies_1988} or axial magnetic flux~\cite{lucibello_2024b}.
It is thought that the flow shear should be comparable to the Alfv\'{e}nic timescale for stability.
Collisionless gyroviscosity phenomena both regulate and produce sheared flows
in the collisionless large Larmor radius regime,
\textit{i.e.}~when the ion inertial length is comparable to the pinch radius and
the ion relaxation time is weakly comparable to the axial flow-through time $\tau = L/v_z$ or other dynamical timescales.

In the collisionless large Larmor radius regime, the ion velocity profile of a strongly forced flow viscously evolves
to be proportional to the magnetic vector potential profile, self-organizing flow shear onto the scale of the pinch radius.
The magnitude of shear is related to the magnitude of pressure anisotropy.
In addition, anisotropy induced by doubly adiabatic flux compression generates radially sheared axial flow.
These phenomena are potentially stabilizing to short-wavelength kink instabilities.
The role of dissipation in the stability of such flows is under investigation.

It is thought that short-wavelength kink modes may be stabilized when the shear frequency $v_z'$
is on the order of the Alfv\'{e}n transit rate, $v_z'=v_A/r_p$.
Considering $v_A/r_p = \alpha_{rz}\omega_c$ leads to an estimate of the characteristic anisotropy associated
with Alfv\'{e}nic shear flow,
\begin{equation}\label{eq:stabilizing_anisotropy}
  \alpha_{rz}^* = \frac{1}{2\sqrt{\mathfrak{B}_i}} = \frac{u_d}{4v_{ti}}  
\end{equation}
where $\mathfrak{B}_i$ is the ion Budker parameter (Eq.~\ref{eq:budker}),
equivalently the inverse drift parameter~\cite{kinetics_i} where $u_d=|u_i-u_e|$.
Thus, the characteristic anisotropy of a sheared-flow pinch in kinetic equilibrium is on the order of its drift parameter.

The ion Budker parameter depends only on the linear density of confined particles.
Hence, a large number of confined particles (implying mostly ion cyclotron orbits, extensively discussed
in the prequel~\cite{kinetics_i}) also requires only a gentle anisotropy for stabilizing
self-organized shear flow in the collisionless regime.
However, such anisotropy becomes hard to generate as the cyclotron orbits discourage agyrotropic anisotropy.
Yet if the magnitude of ion anisotropy is too great then kinetic instabilities will enhance dissipation~\cite{Bott_2024}.
This suggests to target a moderate ion Budker parameter.
To conclude, recent studies have noted the role of flow shear on the ion skin depth scale in regulating dissipation in
astrophysical turbulence studies~\cite{hellinger_2024}.
In the laboratory, the stabilizing effect of sheared flows on the tokamak's internal kink
have been observed to be concurrent with thermal anisotropy~\cite{kink_stabilization}.

\section{Numerical validation of the anisotropy model\label{sec:numerical_validation}}
This section explores the validity of the anisotropy model, and concurrent shear flow generation, described analytically in this work.
This is accomplished via a simple numerical experiment, namely evolving trajectories in a prescribed flux compression and
observing the resulting statistics.
The results support the hypothesis of self-generation of radially sheared axial flow due to doubly adiabatic compression of
ion betatron orbits during ideal flux compression of a large Larmor radius Z pinch.
The numerical experiment is a simple approximation to the plasma kinetics of flux compression, and is not self-consistent
because neither the self-magnetic field of the generated ion flow nor its associated radial electric field are modeled.
Nevertheless, this simple model provides insight into the essential physical models described in this article,
validating the postulated anisotropy profiles and the axial flow generation mechanism.
Self-consistent hybrid particle-in-cell simulations of sheared flow generation during
large Larmor radius Z-pinch compression can be found in Channon et al~\cite{channon_2004}.
A future study of the nonlinear dynamics of collisionless flux compression 
is planned to utilize self-consistent, continuum kinetic Vlasov-Maxwell simulations.

\subsection{Numerical model: evolving trajectories in a compressing flux function}
The collisionless Boltzmann equation is a continuum description of the phase flow generated by an ensemble of discrete trajectories
subject to a certain force field.
A numerical solution of the collisionless Boltzmann equation is provided by sampling a large number of states
from an initial distribution function, evolving the states in a specified field (from either a known solution or from expected dynamics),
and reconstructing the distribution function of the evolved states.
For example, kinetic equilibrium of the Bennett profile may be verified by sampling the initial distribution
(an isothermal drifting Maxwellian with density radially distributed as $n(r)\sim (1+(r/r_p)^2)^{-2}$),
evolving the trajectories in the Bennett field, and verifying stationary statistics of the observables.
The method of observing statistics from trajectory ensembles is fairly common in turbulence studies~\cite{guillevic2023}, and in this
case is applied to observe development of pressure anisotropy and macroscopic flows resulting from flux compression of the Z pinch.

The method consists of the following.
The center-of-charge frame is adopted (cf. the prequel~\cite{kinetics_i}) in which electrons and singly charged ions are
perfectly counterstreaming and the electromagnetic field is entirely magnetic, generated by the two-parameter azimuthal flux per unit length
\begin{equation}\label{eq:az_not_normalized}
  \psi_\theta'(r; I_\infty, r_p) = \frac{\mu_0I_\infty}{4\pi}\ln\Big(1 + \Big(\frac{r}{r_p}\Big)^2\Big)
\end{equation}
where $I_\infty$ is the total current and $r_p$ characteristic radius.
Suppose that Eq.~\ref{eq:az_not_normalized} describes both the initial and final
states of flux compression through the parameter sets $(I_0, r_0)$ and $(I_1, r_1)$, connected by
specified functions of time $I_\infty(t)$ and $r_p(t)$ representing time-varying current and geometry.
The two parameters may be imagined to vary independently~\cite{crews_kadomtsev}, 
relating changes in flux to the current and inductance variations.  
The axial electric field arising from the varying flux is
\begin{equation}\label{eq:axial_electric_field}
  E_z = \frac{\partial\psi_\theta'}{\partial t} = \psi_\theta'\frac{d\ln I_\infty}{dt} - rB_\theta\frac{d\ln r_p}{dt}.
\end{equation}
The two terms on the right-hand side are the parts $\dot{I}L$ and $I\dot{L}$ of the inductive electromotive force $\frac{d}{dt}(IL)$
represented per unit length as the axial electric field. 
Observe that comparing the radial, compressive $\bm{E}\times\bm{B}$ velocity $v_r = \frac{dr}{dt} = -E_z/B_\theta$
to the radial guiding center $r_g=r_g(t)$ of a cyclotron trajectory gives, with $-\frac{E_z}{rB_\theta} = \frac{d\ln r_g}{dt}$,
\begin{equation}\label{eq:geometric_compression}
  \frac{d\ln(r_g/r_p)}{dt} = -\frac{\psi_\theta'}{rB_\theta}\frac{d\ln I_\infty}{dt}.
\end{equation}
Equation~\ref{eq:geometric_compression} suggests that compression is geometrically self-similar
(maintaining the form of Eq.~\ref{eq:az_not_normalized}) if the current is held constant and only inductance is varied.
Thus, for this numerical experiment the parameter $r_p$ is smoothly varied as
\begin{equation}\label{eq:pinch_compression}
  r_p(t) = r_1 + (r_0 - r_1)\frac{1 + \cos\big(\pi\frac{t-t_0}{t_1-t_0}\big)}{2},
\end{equation}
while holding the parameter $I_\infty=I_0$ constant.
The proposed model is a quasi-static approximation which should be valid if the variations
are slow relative to the Alfv\'{e}n transit time.
The similarity of the below results to the self-consistent simulations of Channon et al.~\cite{channon_2004} suggests that
a physical flux transport model is not necessary to validate the doubly adiabatic anisotropy model nor the physical mechanism
of sheared-flow generation.

Ion trajectories sampled from the initial Bennett equilibrium are evolved in the time-varying electric and magnetic fields of
Eqs.~\ref{eq:az_not_normalized} and~\ref{eq:axial_electric_field}.
A radial electric field is not included, meaning that the compression takes place consistently in the
center-of-charge frame and the electric field associated with flow shear is neglected.
The numerical experiment therefore consists of a specified compression of the flux function,
an observed axial acceleration and radial compression of the plasma ions,
and an implied axial acceleration and radial compression of the plasma electrons.

Let the dynamic variables be normalized to the initial thermal state,
\begin{subequations}\label{eq:normalization}
\begin{eqnarray}
  \widetilde{r} &=& \frac{r}{r_0},\\
  \widetilde{\bm{v}} &=& \frac{\bm{v}}{v_0},\\
  \widetilde{t} &=& \frac{t}{r_0/v_0},\\
  \widetilde{\psi}_\theta' &=& \frac{q_i\psi_\theta'}{m_iv_0}
\end{eqnarray}
\end{subequations}
where $q_i, m_i$ are ion charge and mass and $v_0 = \sqrt{kT_0/m_i}$ is the initial thermal velocity.
In the normalized representation, the ion distribution function and the flux function are
\begin{subequations}\label{eq:normalized_distribution_and_flux}
\begin{eqnarray}
  f(\widetilde{r},\widetilde{\bm{v}}) &=& Z_i^{-1}\exp(-\widetilde{\psi}_\theta'/\mathfrak{B}_i)
                                          e^{-\frac{1}{2}(\widetilde{v}_r^2 + \widetilde{v}_\theta^2 + \big(\widetilde{v}_z - \mathfrak{B}_i^{-1/2}\big)^2)}\\
  \widetilde{\psi}_\theta'(\widetilde{r}) &=& 2\mathfrak{B}_i\ln(1 + (\widetilde{r}/\widetilde{r}_p(t))^2)
\end{eqnarray}
\end{subequations}
where $Z_i$ is the partition function and $\mathfrak{B}_i = \frac{q_i^2N}{4\pi\varepsilon_0 m_ic^2}$ is the ion Budker parameter.
The following section presents numerical simulations at varying ion Budker parameter $\mathfrak{B}_i$.
The tildes are now dropped on the normalized variables.

\subsection{Simulated sheared flows at varying orbit magnetization}
This section considers the sheared flows produced by three simulations at ion Budker parameters $\mathfrak{B}_i=(1,10,100)$,
representing the mostly betatron, mixed betatron/cyclotron, and mostly cyclotron regimes.
The simulations are conducted by sampling $10^4$ ions from the initial equilibrium distribution of Eqs.~\ref{eq:normalized_distribution_and_flux}
using the Metropolis algorithm and evolving their trajectories while compressing the scale parameter of the flux function
from $r_0=1$ to $r_1=0.25$ according to Eq.~\ref{eq:pinch_compression} over a period of four normalized (Alfv\'{e}n)
times $\widetilde{T}=4$ (confirmed by experimentation to be sufficiently long to observe adiabatic behavior).
The distribution function is then reconstructed using radial kernel density estimation (KDE)~\cite{radial_kde2023}.
The elementary RK45 time-stepping algorithm of the SciPy \textsc{odeint} routine was found to be sufficient to
reconstruct the relevant statistics given the crudeness of the approximate solution.
A highly resolved, fully self-consistent continuum kinetic simulation of the flux compression process is reserved to future work.

\subsubsection{Anticipated radial anisotropy profiles}
The CGL-like anisotropy model developed in this work for the large Larmor radius Z pinch (Eqs.~\ref{eq:anisotropy_model})
predicts that flux compression induces an approximately radially uniform magnitude of anisotropy $\alpha$ in which
the principal anisotropy axis rotates from field-aligned in the magnetized periphery to current-aligned in the current-dense core.
The anisotropy magnitude is a function of the electric current and geometric compression ratios.
Given the geometric compression ratio $r_1/r_0=0.25$ and the fixed current $I_1/I_0=1$, the anisotropy magnitude
is predicted to be $\alpha = 1-0.25 = 0.75$, and the anisotropy axis should rotate smoothly through the transitional magnetization
layer whose location and thickness, depicted in Fig.~\ref{fig:anisotropy}, is a function of the ion Budker parameter.


\subsubsection{Simulation results}
\begin{figure}[h!]
  \includegraphics[width=0.725\linewidth]{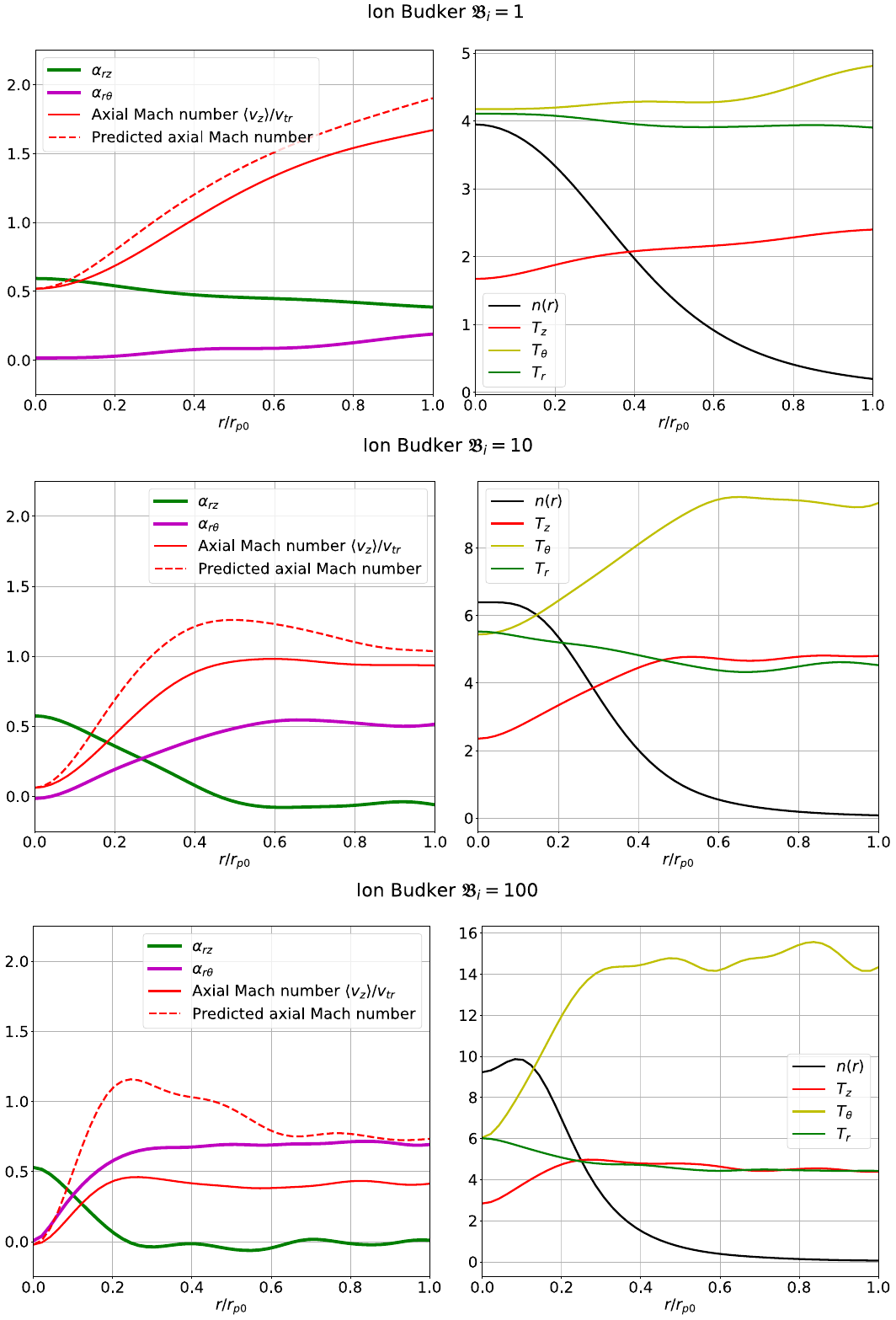}
  \caption{Radial profiles of observables reconstructed after following $N=10^4$
    ion trajectories from their initial kinetic equilibrium throughout an idealized flux compression,
    for a) the fully betatron ($\mathfrak{B}_i=1$),
    b) the mixed betatron/cyclotron ($\mathfrak{B}_i=10$),
    and c) the mostly cyclotron ($\mathfrak{B}_i=100$) trajectory regimes.
    The anisotropy orientation transitioning from agyrotropic in the pinch core to gyrotropic in the periphery 
    agrees with the expected betatron and cyclotron orbital densities (\textit{cf.}~Fig.~\ref{fig:anisotropy}),
    and sheared flows are observed which terminate to a constant velocity
    at the edge of the transitional magnetization layer. 
\label{fig:statistical_profiles}}
\end{figure}

The profiles of temperature, density, anisotropy, and axial Mach number determined from
moments of the estimated distribution function following compression, as obtained by the SciPy \textsc{gaussian\_kde} method,
are plotted in Fig.~\ref{fig:statistical_profiles}.
A predicted axial Mach number is plotted alongside the axial Mach number, based on assuming the axial flow to locally display
the equilibrium condition $v_z'=\alpha_{rz}\omega_{ci}$ and is thus obtainable in quadrature as
\begin{equation}\label{eq:velocity_reconstruction}
  \langle v_z\rangle = v_z|_{r=0} + \int_0^r \alpha_{rz}(r')\omega_{ci}(r')dr'
\end{equation}
where the agyrotropic anisotropy is defined in Eq.~\ref{eq:anisotropy_definitions}.
Agreement of the observed velocity profile with Eq.~\ref{eq:velocity_reconstruction} would suggest the distribution
post-compression is ideally gyromixed into the bi-Maxwellian kinetic equilibrium.
The observed flow is seen to follow the general trend of Eq.~\ref{eq:velocity_reconstruction}
with a disagreement in magnitude, suggesting higher-order corrections in the kinetic equilibrium beyond the simple isothermal
bi-Maxwellian solution.

General features of the solutions depicted in Fig.~\ref{fig:statistical_profiles} are:
\begin{itemize}[topsep=0pt,itemsep=-1ex,partopsep=1ex,parsep=1ex]
\item spatially varying anisotropy profiles $\alpha_{rz}$ and $\alpha_{r\theta}$ which sum approximately to a constant,
  consistent with the anisotropy orientation being proportional to the local density of betatron and
  cyclotron orbits as hypothesized (\textit{cf.}~Fig.~\ref{fig:anisotropy});
\item radially sheared axial flows, depressed in the core and elevated at the edge, and
  terminating at the transition point between the betatron and cyclotron orbital densities;
\item isothermal radial temperature and non-isothermal axial and azimuthal temperatures;
\item depressed axial temperatures associated to radii with mostly betatron orbits,
  and excess azimuthal temperatures associated to radii with mostly cyclotron orbits.
\end{itemize}
It is tempting to associate the depressed axial temperatures around the betatron orbit density to the betatron orbits having greater susceptibility
to axial acceleration by the axial electric field than the cyclotron orbits.
However, observe that the cyclotron orbits at large radii display the fastest axial velocities following compression.
As shown in Section~\ref{sec:spon_sheared_flow}, this occurs because the viscous force induced by anisotropic betatron heating pushes the ion
betatron population in the current-dense core into the $-\hat{z}$-direction, and in the positive $+\hat{z}$-direction throughout the
magnetization transition layer, thus viscously transferring axial momentum to larger radii, accompanied by the emission of sonic waves.
Thus, betatron acceleration in the axial electric field of the compressing azimuthal flux viscously produces faster flow at the edge than in the core.
This counterintuitive result demonstrates the importance of the full spectrum of orbits,
shedding light on the finite-radius process considered by B.~A.~Trubnikov in the singular limit $r_p\to 0$~\cite{trubnikov1992}.

The numerical experiments suggest that higher Budker parameter cases obtain a greater pressure compression ratio.
This discrepancy is a consequence of normalization to the initial states of the simulations.
Although the flux compression ratio is the same for each simulation (evidenced by the consistent anisotropy ratios which develop), the absolute
value of magnetic flux is greater for greater Budker parameters, resulting in higher relative voltages of compression and greater energy transfer.
Note also that the betatron-rich plasmas do not compress geometrically,
which occurs only when the drift motion follows the flux surfaces.

\begin{figure}[h!]
  \includegraphics[width=0.95\linewidth]{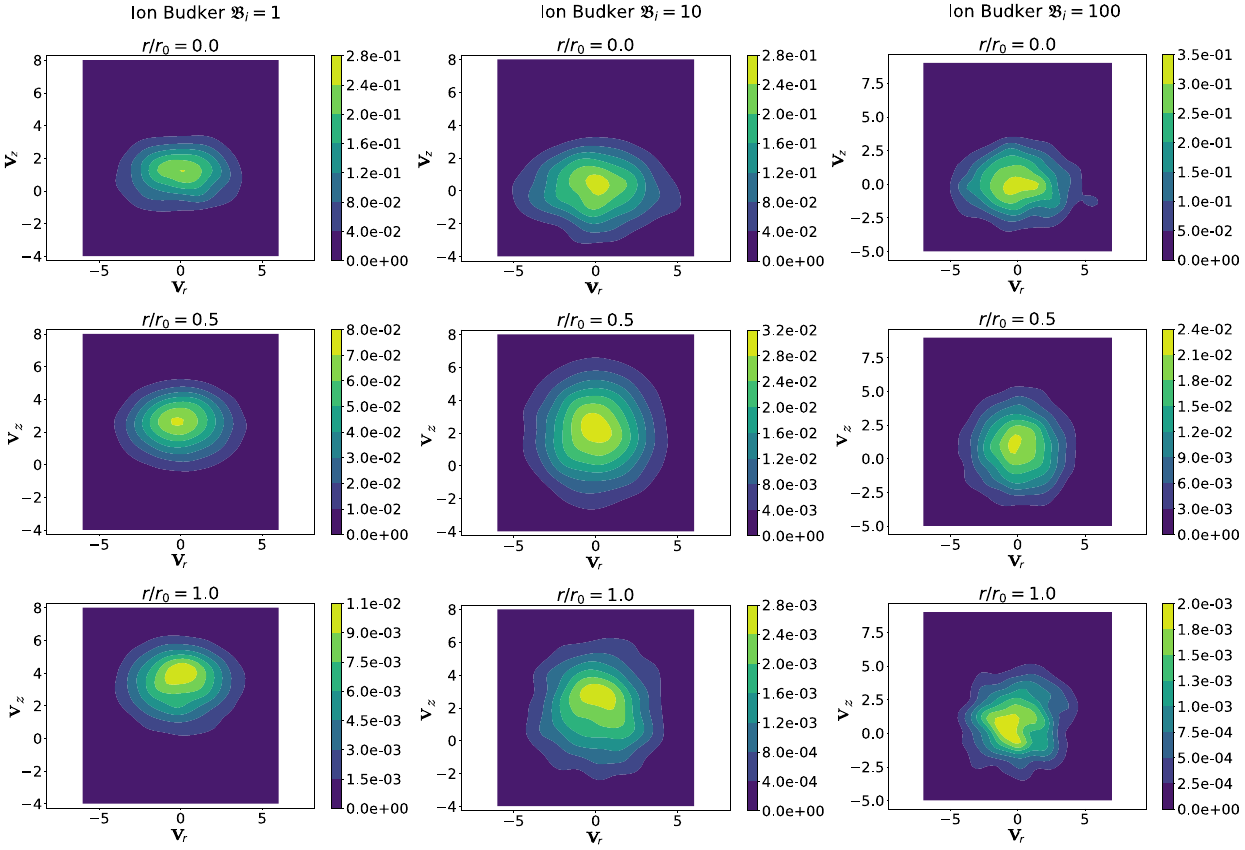}
  \caption{Radial/axial velocity space $(v_r, v_z)$, marginalized on the azimuthal velocity ($\int_{-\infty}^\infty f(r,v_r,v_\theta,v_z)dv_\theta$),
    shown at various radii in the state following compression, from which the moments were obtained for the
    profiles shown in Fig.~\ref{fig:statistical_profiles}.
    The distribution function is well-approximed by a bi-Maxwellian,
    with a skewed shift in the peak indicating higher-order dependence on the flux function,
    as explored in Section~\ref{sec:model-distributions}.
\label{fig:statistical_vspace}}
\end{figure}

Figure~\ref{fig:statistical_vspace} illustrates the reconstructed distribution function following compression
marginalized on the azimuthal velocity ($\int_{-\infty}^\infty f(r,v_r,v_\theta,v_z)dv_\theta$),
providing an alternative perspective on the radial profiles demonstrated in Fig.~\ref{fig:statistical_profiles}.
The full betatron regime ($\mathfrak{B}_i=1$) displays radial heating and an axial dispersion in velocities,
the mixed betatron/cyclotron regime ($\mathfrak{B}_i=10$) shows a more even axial/radial heating with a diminished drift velocity,
and the mainly cyclotron regime ($\mathfrak{B}_i=100$) shows agyrotropic anisotropy only close to the axis.
The distribution functions appear close to an anisotropic Maxwellian, 
indicating that moment models with full pressure tensor dynamics~\cite{srinivasan2018} should be able to observe the
self-generated axial shear flow phenomena explored in this work.

These simulations suggest, within the obvious limits of the non-self-consistent model, that axial flows within the pinch core
may be anticipated during collisionless flux compression of a Z pinch plasma, potentially stabilizing the plasma to the $m=1$ kink instability.
The nature of the self-generated sheared flow profile depends strongly on the distribution of cyclotron and betatron orbits, which is controlled by
the initial linear density of the pinch prior to compression.
An initial deuterium density around $N\approx 10^{19}\text{ m}^{-1}$ ($\mathfrak{B}_i\approx 10$)
appears ideal to generate a sonic sheared flow that terminates just past the pinch radius.
Understanding the self-consistent, gyroviscoresistive evolution of the self-magnetic field due to the induced
ion current and the consequences of microinstabilities triggered by anisotropy remains for future work.

\subsection{The role of thermal anisotropies in radial force balance}
The pressure profiles observed to evolve away from the Bennett profile with the specified flux compression,
which can be understood through analysis of the radial forces induced by the pressure anisotropy.
Specifically, consider a three-temperature Maxwellian distribution of uniform radial temperature $T_r$ but radially varying,
distinct axial and azimuthal temperatures $T_z$ and $T_\theta$,
\begin{equation}\label{eq:three-temperature-maxwellian}
  f_s = Z^{-1}n(r)\exp\Big(-\frac{mv_r^2}{2kT_r} - \frac{mv_\theta^2}{2kT_\theta(r)} - \frac{m(v_z-u_s)^2}{2kT_z(r)}\Big)
\end{equation}
with $Z$ the partition function.
Substitution of Eq.~\ref{eq:three-temperature-maxwellian} into the polar form of the stationary Vlasov equation (Eq.~\ref{eq:polar_vlasov})
and computing the first radial moment yields the pressure balance,
\begin{equation}\label{eq:temperature_gradient_pressure_balance}
  \frac{n'}{n} + \frac{T_\theta'}{T_\theta} + \frac{T_z'}{T_z} + \frac{\alpha_{r\theta}}{r} = \frac{q_s}{m_s kT_r}(E_r + u_sB_\theta)
\end{equation}
where $\alpha_{r\theta} = 1 - T_\theta/T_r$ is the gyrotropic pressure anisotropy.
The terms $T'/T$ are the temperature gradient forces~\cite{Vold_2018}, while the
thermodynamic force $\frac{\alpha_{r\theta}}{r}$ is an apparent force in polar coordinates
describing an imbalance of the Coriolis and centrifugal forces acting on the population within a fluid element.
Given the observed temperature gradients produced by anisotropy, these three thermal forces are directed inwards, which may
flatten and even elevate the density distribution around the axis.
Such features are observed in Fig.~\ref{fig:statistical_profiles}.

\section{Summary of notable results and discussion}\label{sec:conclusion}

Sheared flows are well-recognized elements of high-performance fusion plasmas.
Magnetized plasma viscosity, naturally an important consideration for fusion concepts utilizing sheared flows,
is connected with thermal anisotropies\cite{mahajan_shear}.
In turn, anisotropies are themselves regulated by particle orbits 
through the statistics of adiabatic invariants and kinetic instabilities.
Orbital magnetization and kinetics are thus key considerations for magnetized viscosity.

In the prequel, the spatial distributions of
cyclotron and betatron trajectories were determined for a kinetic pinch equilibrium~\cite{kinetics_i}.
This work applied such concepts to kinetic equilibrium and dynamics.
The anisotropy model of Chew, Goldberger, and Low (CGL), often called doubly adiabatic theory, was extended by taking into account
both the cyclotron and betatron orbits.
The CGL equations were found to accurately describe the magnitude of thermal anisotropy developed in a plasma fluid element, 
yet distinct anisotropy axes were found depending on the proportion of the trajectories.
Specifically, betatron fluids align their anisotropy with the electric current.
Thus, adiabatic variations in the betatron trajectories are associated with agyrotropic anisotropy.

Inviscid flows and collisionless trajectories are usually antonyms;
the collisionless limit is highly viscous. 
On the other hand, collisionless, magnetized sheared flow kinetic equilibria exist 
in which gyro-phase mixing balances with the mixing of nearby trajectories.
Agyrotropic anisotropy is associated with such flows because gyro-phase mixing 
smoothes out variations in the gyro-phase angle.
However, magnetized flows are not inviscid.
It is instructive to consider the initial-value problem of a magnetized flow \textit{out of equilibrium},
in which case phase mixing rapidly brings it \textit{to} equilibrium.
In this sense, the flow is highly viscous, in which gyroviscosity organizes the flow rather than dissipating it entirely.
It is interesting that gyro-phase mixing occurs regardless of whether the underlying
orbits are cyclotrons or betatrons, but the requisite anisotropy is more naturally associated with the betatron orbits.

To understand the consequences of such agyrotropic anisotropies, 
a general class of sheared-flow kinetic pinch equilibrium 
was studied using the method of Hermite polynomials. 
The simplest equilibrium is isothermal and bi-Maxwellian, in which flow is a linear flux function 
and in which flow shear is maximum where local $\beta\approx 1$.

Higher-order equilibria were obtained by an expansion of flow in the vector potential.
Essentially arbitrary equilibrium flows 
are obtained, yet higher-order flux dependence is associated with higher-moment features such as skewness, kurtosis, etc.
Thus, higher-order flux dependence is associated with progressively more non-Maxwellian features,
and in this way the linear flux function flow appears more robust to collisions that higher-order flows.

Kinetic dynamics were then considered in forced shear flows and in response to forced anisotropies.
The former is the forward process of viscosity, in which the flow is phase-mixed to equilibrium, in effect
freezing the velocity profile into a simple flux function.
The latter is an inverse viscous process in which out-of-equilibrium thermal anisotropy self-generates sheared flow,
for example during sufficiently fast adiabatic compression.

The doubly adiabatic model for the betatron fluid was validated
by evolving ion trajectories of an initial Bennett pinch during flux compression.
A radially uniform anisotropy magnitude results, but the 
anisotropy is gyrotropic in the magnetized periphery and agyrotropic in the current-dense core.
Radially sheared axial flows occur through the transitional magnetization layer, in agreement with kinetic equilibrium solutions.

Substantial work remains to be done in Z-pinch kinetic theory.
This article considers only velocity gradients and gyroviscosity;
temperature gradients remain to be studied.  
Beyond equilibrium, the oscillation spectrum and kinetic stability of Z-pinch plasmas  
for Budker parameters in which the ion betatron population cannot be neglected is an open problem.
Some work has been done, valid in the magnetized periphery, utilizing gyrokinetic models.
It is possible that a reduced kinetic description amenable to linear analysis can be developed using the cyclotron-betatron splitting proposed
in the prequel~\cite{kinetics_i}.
This approach is intended to be pursued in future work.

\begin{acknowledgments}
  The authors would like to thank A.E. Robson for valuable discussions on anomalous resistivity that have influenced the development of these ideas,
  A.D. Stepanov for extensive discussions on betatron heating, and J. Coughlin for discussions on gyroviscosity.
  The information, data, or work presented herein is based in part upon work supported by the National Science Foundation under Grant No. PHY-2108419.
\end{acknowledgments}

\section*{Author Declarations}

\subsection*{Conflict of Interest}
The authors have no conflicts to disclose.

\subsection*{Author Contributions}
\textbf{Daniel W. Crews}: Conceptualization (lead), Formal analysis (lead), Investigation (lead), Methodology (lead), Writing - original draft (lead),
Writing - review and editing (lead).
\textbf{Eric T. Meier}: Conceptualization (equal), Project Administration (equal), Supervision (equal), Writing - review \& editing (equal).
\textbf{Uri Shumlak}: Conceptualization (equal), Project Administration (equal), Supervision (equal), Writing - review \& editing (equal).

\subsection*{Data Availability}
The data that support the findings of this study are available from the corresponding author upon reasonable request.


\appendix

\bibliography{shear_flow}

\end{document}